\documentclass[a4paper,11pt]{article}
\usepackage{amsmath}
\numberwithin{equation}{section}
\numberwithin{equation}{subsection}

\usepackage{amssymb}
\usepackage{amsfonts}
\usepackage{color}
\usepackage{graphicx}
\usepackage{wrapfig} %wrapfig.sty is in DIR
\usepackage{slashed}
\usepackage{epic}
\usepackage{eepic}

\def\à{\`a}
\def\ò{\`o}
\def\ì{\`\i}
\def\ù{\`u}
\def\à{\`a}
\def\è{\`e}
\def\é{\'e}
\def\È{\`E}

\begin{document}
\title{{THE CONFORMAL UNIVERSE I: \\
Theoretical Basis of Conformal General Relativity}}
\author{\it{\small FINAL VERSION}\\
    \\
     Renato Nobili\\
    {\normalsize    E-mail: renato.nobili@unipd.it}\\}
\date{Padova, 14 August 2016}
\maketitle
\pagestyle{myheadings} \markright{R.Nobili, The Conformal Universe I}

\begin{abstract}
\noindent This is the first of three papers on Conformal General Relativity (CGR), which differs from Einstein's General
Relativity (GR) in that it requires action--integral invariance under local scale transformations in addition to
general coordinate transformations. The theory is here introduced in the semiclassical approximation as a preliminary approach to a
quantum theoretical implementation. The idea of a conformal--invariant extension of GR was introduced by Weyl in 1919. For
several decades it had little impact, as CGR implies that all fields are massless. Today this does not appear to
be an unsurmountable difficulty since nonzero mass parameters may result from the spontaneous breakdown of conformal symmetry.
The theory leads to very interesting results and predictions: 1) the spontaneous breakdown of conformal symmetry
is only possible in a 4D--spacetime with small negative curvature; 2) CGR requires the introduction of a ghost scalar field
$\sigma(x)$ invested with geometric meaning and a physical scalar field $\varphi(x)$ of zero mass, both of which have
nonzero vacuum expectation values; 3) in order to preserve $S$--matrix unitarity,
$\sigma(x)$ and $\varphi(x)$ must interact in such a way that the total energy density is bounded from below; 4) this interaction
makes $\varphi(x)$ behave like a Higgs field of varying mass, which is capable of promoting a huge energy transfer from geometry
to matter identifiable as the big bang; 5) in the course of time, the Higgs boson mass
becomes a constant and CGR converges to GR.
\end{abstract}

\tableofcontents
\addtocontents{toc}{~\hfill\textbf{Page}\par}
\newpage

\section{Introduction: the problem of the Beginning}
\label{beginning}
In this and two following papers -- Part II and Part III -- a novel theory of inflationary cosmology, based on a
conformal extension of the principle of General Relativity (GR), is presented. In this Part, the theory is
formulated in the semiclassical approximation as the premise of a quantum theoretical implementation.

The need for a revision of the relationship between spacetime and matter is due to the fact that GR suffers
from several critical shortcomings. In particular:
\vspace{-2mm}
\begin{itemize}
\item[1.] The gravitational equation of  Einstein establishes the proportionality of two conservative quantities,
one invested with physical meaning (energy momentum tensor) and the other with geometric meaning (gravitational tensor),
but the former is renormalizable and the latter is not, which is a beautiful contradiction.
\vspace{-2mm}
\item[2.] Even in the semiclassical approximation, GR is unable to provide reasonable solutions to the main problems raised
by modern cosmology. In an attempt to explain the homogeneity and isotropy of the universe on the large scale, and --
more recently -- the anisotropies of cosmic background temperature, the standard model of inflationary cosmology is put in a condition
to invoke the decay of a sort of Higgs boson field, the mass of which is estimated to be about $10^{11}$ times larger than that of
the Higgs boson detected in LHC experiments \cite{MUKHANOV} \cite{LINDE} \cite{LYTH}. This assumption conflicts very seriously with
the Standard Model of elementary particles.
\vspace{-2mm}
\item[3.] GR is unable to explain why the cosmological constant, as vacuum energy--density, is about $10^{-47}$GeV$^4$.
A value as small as this is incompatible with the hypothesis that it is totally or partially due to the zero--point energy
density of all the fields in the quantum vacuum state \cite{WEINBERG0}. Since the contributions to zero--point energy density
are positive for bosons and negative for fermions, and are inversely proportional to positive powers of Planck's constant $h$, we must conclude that the
total zero--point energy density of a possible ``true'' quantum field theory is exactly zero, and such it remains at the classical limit.
This means that, in spite of the non--renormalizability of GR, Bohr's principle correspondence, which states that quantum mechanics
converges to classical mechanics as $h \rightarrow 0$, also holds in quantum field theory, thus legitimating {\em a posteriori}
the semiclassical approximation of the theory.
\vspace{-2mm}
\item[4.] According to Standard Cosmology, going backward in time, we find an initial time at which all spacelike volume--elements
shrink to zero, so that the energy density of matter field becomes infinite, which sounds physically absurd.
\vspace{-1mm}
\end{itemize}

Let us briefly recall the general structure of Einstein's gravitational equation for the purpose of
fixing symbols, sign conventions, measurement units and relevant dimensional constants, and of
introducing the point of departure of my investigation.

Let $g_{\mu\nu}, R_{\mu\nu}, R,\Lambda,\Theta^M_{\mu\nu}$, $G\simeq 6.7086\times 10^{-39}$, $\kappa  \equiv 8\pi G \simeq
1.6861\times 10^{-37}$ GeV$^{-2}$ and $M_{rP} \equiv 1/\sqrt{\kappa} \simeq 2.4354\times 10^{18}$ GeV, be respectively:
{\em metric tensor}, {\em Ricci tensor}, {\em Ricci scalar}, {\em cosmological constant}, {\em EM tensor of matter},
{\em Newton gravitational constant}, {\em (Einstein) gravitational constant} and {\em reduced Planck mass}, and
then consider the structure of Einstein equation $R_{\mu\nu} - \frac{1}{2}g_{\mu\nu}R - g_{\mu\nu}\Lambda =
\kappa\, \Theta^M_{\mu\nu}$. To prevent misunderstandings, let us specify our mathematical conventions:
%\vspace{-2mm}
\begin{itemize}
\vspace{-2mm}
\item[-] The signature of $g_{\mu\nu}(x)$ is $(+,-,-,-)$ and the speed of light is 1.
\vspace{-2mm}
\item[-] $\Theta^M_{\mu\nu}$, which may also depend on $\partial_\lambda g^{\mu\nu}(x)$,  matches Hilbert's definition
\vspace{-2mm}
$$
\Theta^M_{\mu\nu}(x) = 2\,\bigg[\frac{\delta L^M(x)}{\delta
 g^{\mu\nu}(x)}  - \partial_\lambda\frac{\delta L^M(x)}{\delta\,
 \partial_\lambda g^{\mu\nu}(x)} - \frac{g_{\mu\nu}(x)}{2} L^M(x)\bigg]\,,
\vspace{-2mm}
$$
\item[-] $R_{\mu\nu}$ matches Landau--Lifchitz' definition $R_{\mu\nu} = R^\rho_{\,\cdot\,\mu\rho\nu}$
\cite{LANDAU1}, where
\vspace{-2mm}
$$
R^\rho_{\, .\,\mu\sigma\nu}  = \partial_\sigma \Gamma^\rho_{\nu\mu}
-\partial_\nu \Gamma^\rho_{\mu\sigma}+
\Gamma^\lambda_{\mu\nu}\Gamma^\rho_{\lambda\sigma}-
\Gamma^\lambda_{\mu\sigma}\Gamma^\rho_{\lambda\nu}
\vspace{-2mm}
$$
is the Riemann tensor and $\Gamma^\rho_{\nu\mu} = \frac{1}{2} g^{\rho\lambda}\bigl(\partial_\mu g_{\rho\nu} +
\partial_\nu g_{\rho\mu}-\partial_\rho g_{\mu\nu} \bigr)$ are the Christoffel symbols constructed from $g_{\mu\nu}(x)$.
Here and in the following, we put $\partial_\mu f \equiv \partial f/\partial x^\mu$. Since $R^\rho_{\,\cdot\,\mu\sigma\nu}$
is antisymmetric under the interchange $\rho\leftrightarrow \mu$ or $\sigma\leftrightarrow \rho$, the sign of our Ricci
tensor is opposite to that of $R_{\mu\nu} = R^\rho_{\,\cdot\,\mu\nu\rho}$, a definition preferred by other authors
\cite{WEINBERG1} \cite{MISNER}.
\vspace{-2mm}
\item[-] The sign of $\Lambda$ is chosen so that $\rho_{\hbox{\tiny vac}} = \Lambda/\kappa\approx 10^{-47}$GeV$^4$ can be interpreted
as the (positive) energy density of the vacuum. The expression $\Theta^V_{\mu\nu}(x) = \rho_{\hbox{\tiny vac}}\,g_{\mu\nu}(x)$
is here called the {\em EM tensor of the vacuum}.
\vspace{-2mm}
\end{itemize}
With these assumptions and conventions, the gravitational equation of Einstein takes the general form
\vspace{-2mm}
\begin{equation}
\label{Allthetasmunu}
\Theta^M_{\mu\nu}-\frac{1}{\kappa}\Bigl(R_{\mu\nu}- \frac{1}{2}g_{\mu\nu}R\Bigr)
+\rho_{\hbox{\tiny vac}}\,g_{\mu\nu}=0\,.
\vspace{-2mm}
\end{equation}

\newpage
Note that, if we interpret respectively
\vspace{-2mm}
$$
\Theta^G_{\mu\nu}=-\frac{1}{\kappa}\Bigl(R_{\mu\nu}- \frac{1}{2}g_{\mu\nu}R\Bigr)\,, \quad
\Theta^V_{\mu\nu}(x) = \rho_{\hbox{\tiny vac}}\,g_{\mu\nu}\nonumber
\vspace{-2mm}
$$
as the EM tensors of geometry and vacuum, Eq (\ref{Allthetasmunu}) takes the symmetric form
\vspace{-2mm}
\begin{equation}
\label{completegraveq}
\Theta_{\mu\nu}(x)\equiv \Theta^M_{\mu\nu}(x)+ \Theta^G_{\mu\nu}(x) + \Theta^V_{\mu\nu}(x)
= 0\,,
\vspace{-2mm}
\end{equation}
where $\Theta_{\mu\nu}(x)$ represents the {\em total EM tensor of matter, geometry and vacuum}.

Based on these simple considerations, we may speculate that the universe originated from the vacuum state of an
empty world, perhaps as a consequence of a nucleating event capable of priming the spontaneous breakdown of some
fundamental symmetry \cite{COLEMAN1} \cite{COLEMAN2}, and thence evolved under the effect of a huge transfer of
energy from $\Theta^G_{\mu\nu}$ to $\Theta^M_{\mu\nu}$. But the idea of such a prodigious zero--sum
game must be immediately excluded from GR as soon as we realize that the continuity equations
\vspace{-2mm}
\begin{equation}
%\label{separcovcons}
D^\mu\,\Theta^G_{\mu\nu}(x)= 0\,; \quad D^\mu\,\Theta^V_{\mu\nu}(x)=0\,,\quad D^\mu\,\Theta^M_{\mu\nu}(x)= 0\,;
\nonumber
\vspace{-2mm}
\end{equation}
where $D^\mu$ are contravariant spacetime derivatives (see Appendix), hold separately on their own; the first of them
being imposed by the second Bianchi identities \cite{EISENHART}, the second because $D^\lambda g_{\mu\nu}(x)\equiv 0$
holds by definition of $D^\lambda$, as shown by Eqs (\ref{Dmugdownmunu}) and (\ref{Dmugupmunu}) of the Appendix,
and the third by consequence.

No matter how fanciful it may seem, the idea that the big bang may have originated in this way
has the merit of indicating that a suitable modification of GR may account for a non--zero
energy exchange between geometry and matter.

The purpose of this and the next two papers is to prove that this is in fact the case for a theory based on the conformal extension
of GR proposed by Weyl in 1919 \cite{WEYL1} and further developed by Cartan on purely geometric grounds in the early 1920s
\cite{CARTAN1} \cite{CARTAN2} \cite{CARTAN3} \cite{CARTAN4}. In this theory, which I call {\it Conformal General Relativity} (CGR),
a new degree of freedom accounting for a possible scale expansion of spacetime geometry and matter generation
is introduced. We formulate the problem for a spacetime of dimension $n > 2$ (in short, $n$D spacetimes) precisely
to prove that it can be implemented in 4D only.

Among the authors who inspired me with ideas similar to this, I must quote G\"{u}rsey (1963) \cite{GURSEY},
Schwinger (1969) \cite{SCHWINGER}, Fubini (1976) \cite{FUBINI}, Englert {\em et al}. (1976) \cite{ENGLERT1},
Brout {\em et al}. (1978, 1979) \cite{BROUT} \cite{BROUT2}, but my implementation is original and entirely new.

\newpage
\section{Riemann and Cartan manifolds}
Both GR and CGR describe the universe as grounded in a differentiable $n$D manifold parameterized by
adimensional coordinates $x^\mu$ ($\mu =0,1,\dots,n-1$) and enveloped in a continuum of local tangent spaces, all of
which support isomorphic representations of a finite continuous group, called the {\em fundamental group}. The
fundamental group of GR is the $n$D Poincar\'e group and that of CGR is its conformal extension \cite{HAAG}, which is
amply described in \S\,1 of Part II. The structures of both groups can easily be determined according the local
transformation properties of two fundamental tensors, respectively $g_{\mu\nu}(x)$ and $\hat g_{\mu\nu}(x)=
e^{2\alpha(x)}g_{\mu\nu}(x)$, where $e^{2\alpha(x)}$ is the Weyl scale factor.

The local representations of the fundamental group at any two points of the manifold are related to each
other by a generally path--dependent law called the {\em connection}, which describes the
variations in fundamental--group representations as detectable from a local reference frame moved along the path.
Connections which do not return the identity when local frames are moved along closed paths are said
to possess a non--zero {\em curvature}.

In this general scheme, physical fields can be described as irreducible representations of the fundamental
group which, in general, undergo coherent path--dependent transformations when the local frame is moved along
arbitrary manifold paths. The curvature of a connection along an infinitesimal closed path, divided
by the area circumscribed by the path, produces coherent transformations of all field representations, which
may be interpreted as the effects of gravitational forces on the fields grounded in the manifold.

Different groups and types of curvatures characterize different manifolds and manifold connections. Since the fundamental
group of GR preserves the measurement units and line--element lengths, we call the connection {\em metric}
and the spacetime manifold the {\em Riemann manifold}. Since the fundamental group of CGR preserves the angle between
spacetime directions stemming from the same point and in general produces local changes of line--element length, we
call the connection {\em conformal} and the spacetime manifold the {\em Cartan manifold}.

Metric connections are characterized by the property that translations have zero curvature, whereas Lorentz rotations
generally do not. Therefore, in GR, any infinitesimal round--trip of the local reference frame, from a point $x$ to the
same point $x$, generally results in an infinitesimal Lorentz rotation of the frame at $x$. This rotation, which
accounts for the effects of gravitational forces and/or local--frame accelerations, is fully represented by the local
components of a 4--index tensor $R_{\mu\nu\rho\sigma}$, called the {\em Riemann tensor}.

In comparison with metric connections, conformal connections are characterized by possible path--dependent changes of
measurement units. Since we must exclude that round trips may alter the size of a body, dilation connections must have
zero curvature. This means that the dilation connection is the gradient of a scalar field so that the size of a body
is allowed to change only along the time lines of an {\em open spacetime}. In \S\,1 of Part II, we prove that zero curvature
of dilation connection implies zero curvature of all conformal connections with the exception of Lorentz--group connections. Thus,
were it not for the existence of infinitely extended time lines, dilation connections would drive mere gauge
transformations and could therefore be removed from the theory. In this case, CGR and GR would be the same thing.

\subsection{The principle of General Relativity and its significance}
\label{Riemannmanifolds}
Let $L(x) \equiv L\big(\!{\boldsymbol\Phi},\partial_\lambda {\boldsymbol\Phi}, g^{\mu\nu}\!\!, \partial_\lambda g^{\mu\nu}\!\big)$
be the total Lagrangian density of a field theory grounded in a Riemann manifold parameterized by $n>2$  adimensional spacetime coordinates
$x^\mu$ ($\mu =0, 1, \dots ,n-1$) equipped with metric tensor $g_{\mu\nu}(x)$  and volume element $\sqrt{-g(x)}\,d^nx $,
where $g(x)$ is the determinant of matrix $\bigl[g_{\mu\nu}(x)\bigr]$. $L(x)$ is a function of a set of fields ${\boldsymbol\Phi}(x)=
\{\Phi_1(x),\dots, \Phi_m(x)\}$, metric tensor $g^{\mu\nu}(x)$, where $g^{\mu\nu}(x)$ are defined by $g_{\mu\lambda}(x)\, g^{\lambda\nu}(x)= \delta^\nu_\mu$, and their partial derivatives $\partial_\lambda {\boldsymbol\Phi}(x)$, $\partial_\lambda g^{\mu\nu}(x)$.
For the sake of simplicity, we exclude fermions from ${\boldsymbol\Phi}$, since otherwise we should represent the metric by
{\em soldering forms} $e^I_\mu(x)$, $(I=1,\dots n)$ (called {\em tetrads} in 4D), rather than $g_{\mu\nu}(x)$. This is not so
restrictive because, using  $e^I_\mu$ instead of $g_{\mu\nu}$, would give the same result. Thus, the total action integral is
\vspace{-2mm}
\begin{equation}
\label{ATOT}
A  =  \int\sqrt{-g}\,L\big({\boldsymbol\Phi},\partial_\lambda\!{\boldsymbol\Phi}, g^{\mu\nu}\!,\,
\partial_\lambda g^{\mu\nu}\big)\,d^nx \,.
\vspace{-1mm}
\end{equation}

The {\em principle of General Relativity} states that the gravitational equation is determined by the
condition of $A$ invariance under general coordinate transformations, and we must also require the latter
to be continuous and differentiable everywhere, i.e., belonging to the infinite group ${\cal D}$ of
{\em coordinate diffeomorphisms}. Let us see how the condition works.

To fix the notation, let us represent the action of a diffeomorphism $D\in {\cal D}$ on $x^\mu$ as $D\!\!:
x^\mu \rightarrow \bar x^\mu(x)$, and that on a scalar function $f(x)$ as $D\!\!: f(x) \rightarrow \bar f(\bar x)
=f[x(\bar x)]$. Since the squared line--element $ds^2 = g_{\mu\nu}(x)\,dx^\mu dx^\nu$ and the volume element $\sqrt{-g(x)}\,d^n x$ of the manifold are scalar functions of $x$, $D$ acts on these as follows
\vspace{-3mm}
\begin{equation}
\label{gmunu2bargmunu}
D\!\!: ds^2 = g_{\mu\nu}(x)dx^\mu dx^\nu\!\rightarrow\!d\bar s^2= \bar g_{\mu\nu}(\bar x)d\bar x^\mu d\bar
x^\nu;\quad D\!\!: \sqrt{-g(x)}d^n x\!\rightarrow\!\sqrt{-\bar g(\bar x)}d^n\bar x,
\vspace{-1mm}
\end{equation}
where $\bar g_{\mu\nu}(\bar x)\equiv g_{\rho\sigma}[x(\bar x)]$ and $\bar g(\bar x)\equiv g[x(\bar x)]$; and, in general, we have $ds^2 \neq d\bar s^2$.

We now consider the following infinitesimal diffeomorphism $D_\varepsilon \in {\cal D}$ of parameters $x^\mu$
\begin{equation}
\vspace{-2mm}
\label{deltaepsxmu}
D_\varepsilon\!\!: x^\mu(x)\rightarrow \bar x^\mu(\bar x)= x^\mu + \delta_\varepsilon x^\mu \equiv x^\mu + \varepsilon^\mu(x),
\end{equation}
with $\varepsilon^\mu(x)$ infinitesimal, differentiable functions of $x^\nu$. We therefore have the matrices
\vspace{-2mm}
\begin{equation}
\label{parbarxparx}
\frac{\partial \bar x^\mu}{\partial x^\nu} = \delta^\mu_\nu +  \partial_\nu\varepsilon^\mu(x)\,;
\quad \frac{\partial x^\mu}{\partial \bar x^\nu} = \delta^\mu_\nu -  \partial_\nu\varepsilon^\mu(x)\,,
\vspace{-2mm}
\end{equation}
the second of which is clearly the inverse of the first.
Let $g_{\rho\sigma}(x)$ and $\bar g_{\mu\nu}(\bar x)$ be the metric tensors as functions of $x^\mu $ and of $\bar x^\mu$, respectively. From the first of Eqs (\ref{gmunu2bargmunu}) and the second of Eqs (\ref{parbarxparx}),
we derive the infinitesimal variation of $ds^2(x)$
\vspace{-2mm}
\begin{equation}
\label{deltagmunuofx}
D_\varepsilon ds^2(x)= \big[\varepsilon^\lambda(x)\,\partial_\lambda g_{\mu\nu}(x)  -g_{\mu\lambda}(x)\,\partial_\nu\varepsilon^\lambda(x) -g_{\nu\lambda}(x)\,\partial_\mu\varepsilon^\lambda(x)\big]dx^\mu dx^\nu.
\vspace{-2mm}
\end{equation}

Using $g^{\mu\lambda}(x)\,g_{\lambda\nu}(x) = \delta^\mu_\nu$, where $\delta^\mu_\nu$ is Kronecker delta,
we obtain
\vspace{-2mm}
\begin{eqnarray}
\label{deltagupmunu}
D_\varepsilon g^{\mu\nu}(x) = - g^{\mu\rho}(x)\,g^{\nu\sigma}(x)\,D_\varepsilon
g_{\rho\sigma}(x) \,.
\vspace{-2mm}
\end{eqnarray}
Also, using the well--known differential formula $d\,\hbox{det}(M) = \hbox{det}(M)\,\hbox{Tr}[M^{-1}dM]$,
where $M$ is any square matrix and Tr$[\dots]$ stands for matrix trace, we obtain
\vspace{-2mm}
\begin{equation}
\label{deltasrt-g}
D_\epsilon \sqrt{-g(x)} = -\frac{\sqrt{-g(x)}}{2}\,g_{\mu\nu}(x)\,D_\varepsilon g^{\mu\nu}(x)\,.
\vspace{-2mm}
\end{equation}
Lastly, as we can easily verify, under the action of $D_\varepsilon$, $g^{\mu\nu}(x)$
undergo the variation stated by Eq (\ref{deltagupmunu}) and $\Phi_i(x)$ and their
partial derivatives,  variations $\Phi_i(x) = \varepsilon^\mu(x)\,\partial_\mu \Phi_i(x)$ and
$\delta_\varepsilon \partial_\nu \Phi_i(x) = \varepsilon^\mu(x)\,\partial_\mu \partial_\nu \Phi_i(x)$,
respectively.

The general principle of Einstein's relativity requires the variation of action--integral $A$ under arbitrary infinitesimal
diffeomorphisms $D_\varepsilon$
to be zero, i.e.,
\vspace{-1mm}
\begin{equation}
\label{deltaepsA00}
D_\varepsilon A\equiv
\int \frac{\delta A}{\delta g^{\mu\nu}(x)}\,D_\varepsilon g^{\mu\nu}(x)\,d^nx = 0.
\vspace{-1mm}
\end{equation}

Carrying out the variations of Lagrangian density, and then exploiting Eq (\ref{deltasrt-g}), we obtain
\vspace{-2mm}
\begin{eqnarray}
\label{deltaAM0}
\hspace{-4mm}
\frac{\delta  A}{\delta g^{\mu\nu}(x)}  & = & \frac{\delta \big[\sqrt{-g(x)}\, L(x)\big]}{\delta g^{\mu\nu}(x)} -\partial_\lambda
\frac{\delta\big[\sqrt{-g(x)}\,L(x)\big]}{\delta\big[\partial_\lambda g^{\mu\nu}(x)\big]}\equiv \nonumber\\
& & \sqrt{-g(x)}\,\bigg\{\frac{\delta L(x)}{\delta g^{\mu\nu}(x)} -\partial_\lambda
\frac{\delta L(x)}{\delta\big[\partial_\lambda g^{\mu\nu}(x)\big]}- \frac{g_{\mu\nu}(x)}{2}\, L(x) \bigg\}=0\,.
\vspace{-2mm}
\end{eqnarray}

As Eq (\ref{deltaAM0}) has the general form of a motion equation for a field, we can legitimately regard $g_{\mu\nu}(x)$
as the gravitational field (possibly up to a constant metric--tensor term). Since, according to Hilbert's definition,
total EM tensor $\Theta_{\mu\nu}(x)$  is related to $A$ by equation
$$
\Theta_{\mu\nu}(x) = \frac{2}{\sqrt{-g(x)}}\,\frac{\delta  A}{\delta g^{\mu\nu}(x)}\,,
$$
we see that the gravitational equation can be simply written as $\Theta_{\mu\nu}(x)=0$.

Now, let us decompose the total action integral in the form $A=A^M+A^G+A^V$, where
\vspace{-3mm}
\begin{eqnarray}
\label{AM00}
A^M & = & \int\sqrt{-g}\,L^M\big({\boldsymbol\Phi}, \partial_\lambda\! {\boldsymbol\Phi},\,
g^{\mu\nu} \big)\,d^nx \,;\\
\vspace{-2mm}
\label{AG00}
A^G  & = & -\frac{1}{2\kappa} \int\sqrt{-g}\,R\big(g^{\mu\nu},\partial_\lambda g^{\mu\nu}\big)\,d^nx\,;\\
\vspace{-2mm}
\label{AV00}
A^V & = & -\rho_{\hbox{\tiny vac}}\int\sqrt{-g}\,d^nx \,;
\vspace{-2mm}
\end{eqnarray}
are the action integrals respectively of matter, vacuum and geometry. Here, $L^M$ is the Lagrangian density of the interaction
of $g^{\mu\nu}$ with a set of matter fields ${\boldsymbol\Phi}$,  $R$ is the Ricci scalar (with the sign convention prescribed in
\S\,\ref{beginning}) as a function of $g^{\mu\nu}(x)$ and $\partial_\lambda g^{\mu\nu}(x)$, and  $\rho_{\hbox{\tiny vac}}$ the
cosmological constant as energy density of the vacuum.

Therefore, Eq (\ref{deltaAM0}) yields the following variations of $A^M, A^G$ and $A^V$ with respect to the components of
the gravitational field
%\vspace{-2mm}
\begin{eqnarray}
\label{deltaAM00}
& & \hspace{-14mm}\frac{2}{\sqrt{-g}}\bigg[\frac{\delta A^M}{\delta g^{\mu\nu}}-
\partial_\lambda\Big(\frac{\delta\, A^M}{\delta\partial_\lambda g^{\mu\nu}}\Big)\bigg]=
2\bigg[\frac{\partial L^M}{\delta\ g^{\mu\nu}} -\partial_\lambda\Big(\frac{\delta L^M}
{\delta\partial_\lambda g^{\mu\nu}}\Big) - \frac{g_{\mu\nu} L^M}{2}\bigg]= \Theta^M_{\mu\nu}\,,\\
\label{deltaAG00}
& & \hspace{-14mm}\frac{2}{\sqrt{-g}}\bigg[\frac{\delta A^G}{\delta g^{\mu\nu}}-
\partial_\lambda\Big(\frac{\delta A^G}{\delta\partial_\lambda g^{\mu\nu}}\Big)\bigg] =
-\frac{1}{\kappa}\Bigl(R_{\mu\nu} -\frac{1}{2}g_{\mu\nu}R \Bigr)=\Theta^G_{\mu\nu}\,,\\
\label{deltaAV00}
& & \hspace{-14mm}\frac{2}{\sqrt{-g}}\frac{\delta\ A^V}{\delta g^{\mu\nu}}=
\rho_{\hbox{\tiny vac}}\,g_{\mu\nu}=\Theta^V_{\mu\nu}\,,
%\vspace{-2mm}
\end{eqnarray}
so that we have $\Theta_{\mu\nu}= \Theta^M_{\mu\nu} + \Theta^G_{\mu\nu}+ \Theta^V_{\mu\nu} = 0$,
as anticipated in \S\,\ref{beginning}.

Particularly important is the subgroup ${\cal D}^*_\varepsilon\subset {\cal D}_\varepsilon$ of infinitesimal diffeomorphisms
which leave invariant infinitesimal scalar products $(dx_1, dx_2) = g_{\mu\nu}(x)\, dx^\mu_1 dx^\nu_2$.
These represent the condition for $D_\varepsilon$ to leave invariant $ds^2$ and the angle of any two curves passing through
a point of the manifold. Therefore, for all  $D^*_\varepsilon\in {\cal D}^*_\varepsilon$, so that $D^*_\varepsilon\!\!: x^\mu
\rightarrow \bar x^\mu  =x^\mu+{\varepsilon^*}^\mu(x)$, we have
$D^*_\varepsilon ds^2(x) = \bar g_{\mu\nu}(\bar x)d\bar x^\mu_1 d\bar
x^\nu_2 -g_{\mu\nu}(x)dx^\mu_1 dx^\nu_2 =0$; and therefore, from Eq (\ref{deltagmunuofx}),
\vspace{-1mm}
\begin{equation}
\label{bargmunubarx}
\varepsilon^\lambda(x)\,\partial_\lambda g_{\mu\nu}(x)  -g_{\mu\lambda}(x)\,\partial_\nu\varepsilon^\lambda(x) -g_{\nu\lambda}(x)\,\partial_\mu\varepsilon^\lambda(x)=0\,.
\vspace{-1mm}
\end{equation}

Thus, since $D^*_\varepsilon$ does not change $ds(x)$,
$\bar g_{\mu\nu}(\bar x)$ can be regarded as a slightly different parameterization of $g_{\mu\nu}(x)$. In other terms, Riemann manifolds
which are related by $D^*_\varepsilon$ represent the same manifold; this allows us to regard the manifold as an objective entity.
Since ${\cal D}^*_\varepsilon$ is the analog of the group of infinitesimal gauge transformations in electrodynamics,
it is called the infinitesimal {\em gauge group of GR}.

To find the conditions for ${\varepsilon^*}^\lambda(x)$ to generate an infinitesimal gauge transformation, it is suitable
to use the fundamental properties of covariant derivatives $D_\mu$ stated in the Appendix by Eqs (\ref{Dmugdownmunu})
(\ref{Dmugupmunu}), which we rearrange here in a different indexing as
\vspace{-1mm}
\begin{equation}
\label{COVDER}
D_\lambda g_{\mu\nu} \equiv \partial_\lambda g_{\mu\nu}-
g_{\rho\mu} \Gamma_{\lambda\nu}^\rho  - g_{\rho\nu}\Gamma_{\lambda\mu}^\rho = 0\,,\quad
D_\lambda g^{\mu\nu} \equiv \partial_\lambda g^{\mu\nu}+ g^{\mu\rho} \Gamma_{\lambda\rho}^\nu + g^{\rho\nu}\Gamma_{\lambda\rho}^\mu = 0\,,
\vspace{-2mm}
\end{equation}
where $\Gamma_{\mu\nu}^\lambda(x)$ are Christoffel symbols.

From Eq (\ref{bargmunubarx}) and the first of Eqs (\ref{COVDER}),
we obtain
\vspace{-1mm}
\begin{eqnarray}
& & g_{\mu\lambda}\,\partial_\nu {\varepsilon^*}^\lambda
+g_{\nu\lambda}\,\partial_\mu{\varepsilon^*}^\lambda
- {\varepsilon^*}^\lambda\partial_\lambda g_{\mu\nu}  = \nonumber \\
\label{covgaugeid}
& & g_{\mu\lambda}\,\partial_\nu{\varepsilon^*}^\lambda+g_{\nu\lambda}\,\partial_\mu{\varepsilon^*}^\lambda
+g_{\rho\mu}\Gamma_{\lambda\nu}^\rho{\varepsilon^*}^\lambda + g_{\rho\nu}\,\Gamma_{\lambda\mu}^\rho {\varepsilon^*}^\lambda   =0\,,
\vspace{-1mm}
\end{eqnarray}
from which we derive $g_{\mu\lambda}\,\big(\partial_\nu{\varepsilon^*}^\lambda + \Gamma_{\nu\rho}^\lambda\,{\varepsilon^*}^\rho\big)
+ g_{\nu\lambda}\,\big(\partial_\mu{\varepsilon^*}^\lambda + \Gamma_{\mu\rho}^\lambda\,{\varepsilon^*}^\rho\big)= 0$, or,
because of the ``transparency'' property $g_{\mu\lambda}D_\nu X= D_\nu g_{\mu\lambda}X$ described in the Appendix, the
conditions
\vspace{-1mm}
\begin{equation}
\label{GRgauge1}
g_{\mu\lambda}D_\nu {\varepsilon^*}^\lambda + g_{\nu\lambda}D_\mu{\varepsilon^*}^\lambda =D_\mu{\varepsilon^*}_\nu +
D_\nu{\varepsilon^*}_\mu =0\,.
\vspace{-1mm}
\end{equation}

In a similar way, from Eq (\ref{bargmunubarx}) and the second of Eqs (\ref{COVDER}),
we obtain
\vspace{-1mm}
\begin{eqnarray}
&& g^{\mu\rho}\, \partial_\rho{\varepsilon^*}^\nu +g^{\nu\rho}\,\partial_\rho{\varepsilon^*}^\mu - g^{\mu\rho}g^{\nu\sigma}{\varepsilon^*}^\lambda\,\partial_\lambda g_{\rho\sigma}= \nonumber\\
\label{contrgaugeid}
\vspace{-1mm}
&& g^{\mu\rho}\,\partial_\rho{\varepsilon^*}^\nu +  g^{\nu\rho}\partial_\rho{\varepsilon^*}^\mu  + g^{\mu\rho} \Gamma_{\lambda\rho}^\nu{\varepsilon^*}^\lambda +
g^{\nu\rho}\Gamma_{\lambda\rho}^\mu{\varepsilon^*}^\lambda=0\,,
\vspace{-1mm}
\end{eqnarray}
where identities $g^{\rho\nu}\partial_\lambda g_{\mu\rho} = -g_{\mu\rho} \partial_\lambda g^{\rho\nu}$, $g^{\mu\rho}\,g^{\nu\sigma}\Gamma_{\lambda\sigma}^\tau g_{\rho\tau} = g^{\nu\sigma}\Gamma_{\lambda\sigma}^\mu$,
$g^{\mu\rho}\,g^{\nu\sigma}\Gamma_{\lambda\rho}^\tau g_{\sigma\tau}= g^{\mu\rho} \Gamma_{\lambda\rho}^\nu$, and the second of Eq (\ref{covgaugeid}), have been used. Thus, putting $\Gamma^{\mu\nu}_\lambda \equiv g^{\rho\nu}\Gamma_{\rho\lambda}^\mu$,  we obtain
%\vspace{-3mm}
\begin{equation}
\label{GRgauge2}
\partial^\mu{\varepsilon^*}^\nu + \Gamma^{\mu\nu}_\lambda{\varepsilon^*}^\lambda + \partial^\nu{\varepsilon^*}^\mu +
\Gamma^{\mu\nu}_\lambda{\varepsilon^*}^\lambda =  D^\mu{\varepsilon^*}^\nu + D^\nu{\varepsilon^*}^\mu =0,
\end{equation}
Similar conditions for ${\varepsilon^*}^\lambda(x)$ are equivalently described by Eqs (\ref{GRgauge1}) and (\ref{GRgauge2}).

Since gauge transformations leave invariant the objective structure of the Riemann manifold, they also leave invariant each
term of the action integral, independently of each other. This means that equation $D^*_\varepsilon A=0$ does not provide any link among
the different component of $\Theta_{\mu\nu}(x)$, and are therefore ineffective in producing the gravitational equation.
We must therefore reformulate the principle of GR, by stating the invariance of $A$ with respect to coordinate transformations
which belong to factor group ${\cal D}_\varepsilon/{\cal D}^*_\varepsilon$.

\subsection{The principle of Conformal General Relativity and its significance}
\label{Cartanmanifolds}
The fundamental group of a Cartan--manifold in $n$D is the conformal extension of the $n$D Poincar\'e group
which, for the sake of brevity, will be called the $n$D {\em conformal group}. It includes the subgroups of
{\em dilations} and {\em special conformal transformations}, or {\em elations} (term coined by Cartan in 1922),
as amply described in \S\,1 of Part II.

Since the group of Riemann--manifold diffeomorphisms $\cal D$ includes conformal diffeomorphisms,
the {\em principle of CGR} is the same as that of GR. But it differs from the latter in that the
gauge group of CGR is invariant under {\em conformal gauge diffeomorphisms} $C^*$ rather than {\em metric
gauge diffeomorphisms} $D^*$.

The gauge group of conformal diffeomorphisms ${\cal C}^*$ is obtained by combining the gauge group ${\cal D}^*$  of
metric diffeomorphisms described in the previous subsection, with the group of {\em Weyl transformations},
which are defined by the following action on the line element of the Riemann manifold
\begin{equation}
\label{Weyltransf0} ds(x)\rightarrow d\bar s(x) = e^{\beta(x)}ds(x)\,,
\end{equation}
where $\beta(x)$ is a real differentiable function of $x^\mu$. The exponential form of factor $e^{\beta(x)}$ is intended to
ensure its positivity. Correspondingly, metric tensor $g_{\mu\nu}(x)$ and volume element $\sqrt{-g(x)}\,d^nx$ undergo
the following Weyl transformations
\begin{eqnarray}
\label{Weyltransf1} & & g_{\mu\nu}(x)\rightarrow \bar g_{\mu\nu}(x) =
e^{2\beta(x)}g_{\mu\nu}(x)\,;\quad g^{\mu\nu}(x)\rightarrow
\bar g^{\mu\nu}(x) = e^{-2\beta(x)}g^{\mu\nu}(x)\,;\\
\label{Weyltransf2} & & \sqrt{-g(x)}\,d^nx \rightarrow
\sqrt{-\bar g(x)}\,d^nx=e^{n\beta(x)}\sqrt{-g(x)}\,d^nx\,.
\end{eqnarray}
As a bonus, this inclusion generates local elations automatically (Haag, 92).

By combining a gauge metric diffeomorphism $D^*\in {\cal D}^*$ with a Weyl transformation,
we obtain a {\em conformal gauge diffeomorphism} $C^*\in {\cal C}^*$, which acts on $g_{\mu\nu}(x)$ and  $\sqrt{-g(x)}$ as follows
\begin{eqnarray}
\label{gmunu2tildegmunu}
&&C^*\!\!:g_{\mu\nu}(x) \rightarrow \bar g_{\mu\nu}(\bar x) = e^{2\bar\beta(\bar x)}\bar g_{\rho\sigma}(\bar x)\,
\frac{d x^\rho}{d\bar x^\mu}\frac{d x^\sigma}{d\bar x^\nu}\,,\\
\label{tilsqrtdetgmunu}
&&C^*\!\!:\sqrt{- g(x)} \rightarrow \sqrt{-\bar g(\bar x)}=e^{n\bar\beta(\bar x)} \sqrt{-\bar g(\bar x)}
\,\frac{d^n\bar x}{\,d^n x}\,.
%\sqrt{- g(x)}& \rightarrow &\sqrt{-\bar g(\bar x)}=e^{n\bar\beta$\beta(\bar x)} \sqrt{-g[x(\bar x)]}\,,
\end{eqnarray}
Here, $x(\bar x)$ means $x$ as a function of $\bar x$, and we may therefore write $\bar\beta(\bar x)\equiv \beta[x(\bar x)]$,
$\bar g_{\mu\nu}(\bar x) \equiv g_{\mu\nu}[x(x)]$, $\bar g(\bar x)\equiv g[x(\bar x)]$ and, in general,
$\bar f(\bar x)\equiv f[x(\bar x)]$.

The {\em first basic assumption} of CGR is that the fundamental tensor of the Cartan manifold has the general form
\begin{equation}
\label{fundtensor}
\hat g_{\mu\nu}(x) = e^{2\alpha(x)}g_{\mu\nu}(x)\,,
\end{equation}
where $g_{\mu\nu}(x)$ is a pseudo--Riemannian metric tensor and $e^{\alpha(x)}$ is a geometric degree of
freedom accounting for possible local changes of spacetime scale (not a gauge), both  of which are
presumed to depend on the details of matter distribution and its dynamics. In particular, $g_{\mu\nu}(x)$
is presumed to contain information about the gravitational forces locally acting on the matter field,
and $e^{2\alpha(x)}$ information about possible dilation of geometry and matter fields
during the inflationary epoch.

To clarify the significance of this factorization, note that, by performing conformal diffeomorphism $\hat g_{\mu\nu}(x)
\rightarrow g_{\mu\nu}(x) =\hat g_{\mu\nu}(x)/\sqrt{-\hat g(x)}$, we obtain $\sqrt{-g(x)}=1$ and $\sqrt{-\hat g(x)}=
e^{n\alpha(x)}$. A metric tensor of this form is called {\em unimodular}. Therefore, provided that the metric diffeomorphisms
of coordinate parameters are restricted to the subgroup of coordinate diffeomorphisms which preserve unimodularity, one
degree of freedom of the fundamental tensor can be permanently transferred from $\hat g_{\mu\nu}(x)$
to the global scale factor $e^{\alpha(x)}$.

For consistency with the idea of inflation as a process driven by $e^{\alpha(x)}$, $\alpha(x)$ must depend on $x$ through
a time--like function $\tau(x)$ which defines a family of iso--dilation surfaces $\tau(x)=$ const. If this were not so, the
dilation connection of the Cartan manifold could not be represented as the gradient of a scalar quantity. The only way to
implement this property is that to make $\tau(x)$ the length of the worldline segment stemming from the origin $x=0$ of a future
light--cone and ending at a point $x$ of the cone interior, implying that the Cartan manifold itself is entirely confined
to the interior of the future--cone. Clearly, to satisfy this property, the factorization of fundamental tensor
(\ref{fundtensor}) into a global and a local part cannot be arbitrary. Further details are given in \S\,\ref{synchrobs}.

The {\em second basic assumption} is that CGR should rapidly converge to GR at the end of the inflationary epoch, because we
know that, long after the inflation epoch, the geometry of the universe is that of GR or one very close
to it. To be consistent with this assumption, we postulate that $e^{\alpha(x)}$ was very small at the beginning
of the inflation epoch, then rapidly increasing and converging to one during the transition to the post--inflationary era.
In other terms, $\alpha(x)$ must initially be negative, then it must increase and converge to zero at the end
of inflationary epoch. In order for this to occur, the action integral of CGR must have a suitable structure, which is in fact possible,
as we shall see in the following sections. It is thus clear that the preliminary question about CGR is
whether these properties can be effectively implemented in a suitable field theory. The main result of the study
reported in this Part I is that the answer to this question is not only positive, but also quantitatively
satisfying, provided that the matter primarily generated by the inflation process is a Higgs--boson field.
Further details are given in \S\,\ref{G2MneedsH}.

The {\em third basic assumption} of CGR is that unimodular metric tensor $g_{\mu\nu}$ has the
form
\begin{equation}
\label{detgmunu}
g_{\mu\nu}(x)= \gamma_{\mu\nu}(x)+ h_{\mu\nu}(x)\,,\quad \mbox{with }\,\,
\mbox{det}[\gamma_{\mu\nu}(x)] = \mbox{det}[\gamma_{\mu\nu}(x)+ h_{\mu\nu}(x)]=1\,.
\end{equation}
Here, $\gamma_{\mu\nu}(x)$ is the metric tensor representing the average geometry of the universe as determined by the overall
distribution of matter (see \S\,\ref{confbackgrnd}), and $h_{\mu\nu}(x)$ represents the gravitational field as a deviation
from $\gamma_{\mu\nu}(x)$ determined by the local details of matter distribution. If $h_{\mu\nu}(x)$ are so small that they
can be regarded as an infinitesimal perturbation of $\gamma_{\mu\nu}(x)$, the second part of Eq (\ref{detgmunu}) implies
$g^{\mu\nu}(x)\, h_{\mu\nu}(x) \simeq \gamma^{\mu\nu}(x)\, h_{\mu\nu}(x)=0$.

\subsection{Conformal General Relativity on a Cartan manifold}
\label{CGRonCM}
The convergence of CGR to GR at the post--inflationary limit suggests a method for inferring
CGR action integrals from GR action integrals. Remaining unspecified in this inference
are the properties that a GR action integral should have in order to satisfy the basic assumptions about
CGR discussed in the previous subsection. In this subsection, we limit ourselves to describing
the method for the sole purpose of showing in advance some conceptual difficulties that the reader
may encounter in dealing with the mathematical formalism of CGR.

Henceforth, to distinguish the mathematical formalisms of GR from that of CGR, we adopt the following
convention: {\em all quantities related to or grounded in the Cartan manifold are superscribed by a hat}.

The prominent characteristic of Cartan manifold representations is that scale factor $e^{\alpha(x)}$
can be interpreted as the physical promoter of spacetime--scale expansion and matter generation. This is possible
in CGR because each physical field $\Psi$ of GR is replaced by its {\em inflated} counterpart $\hat\Psi(x)$,
defined by the Weyl transformation
\begin{equation}
\label{Psi2tildePsi}
\Psi(x)\rightarrow \hat\Psi(x) = e^{w_\Psi\,\alpha(x)}\Psi(x)\,,
\end{equation}
where $w_\Psi$ is the {\em dimension}, or {\em weight}, of  $\Psi$, as if during inflation
all fields were subjected to a sort of continuous renormalization. The transition from a Riemann
to a Cartan manifold deeply alters the structure of action integrals, partly
because all basic quantities and operators of standard tensor calculus -- namely, manifold
parameters $x^\mu$, partial derivatives $\partial_\mu \equiv \partial/\partial x^\mu$, metric tensor
$g_{\mu\nu}$, covariant derivatives $D_\mu$, Christoffel symbols $\Gamma^\lambda_{\mu\nu}$,
Ricci tensor $R_{\mu\nu}$, Ricci scalar $R$, etc. -- are respectively replaced by their conformal
(tilde) counterparts. These depend on both the standard tensors of GR and the scale factor
$e^{\alpha(x)}$, as described by Eqs (\ref{Gammavariations}) to (\ref{tildeDvT2}) and hereafter
briefly listed:
\begin{eqnarray}
%&& dx^\mu \rightarrow d\tilde x^\mu =e^{\alpha(x)} d x^\mu\,;\quad x^\mu \rightarrow \tilde x^\mu\,;
%\quad\partial_\mu \rightarrow \tilde \partial_\mu = e^{-\alpha(x)} \partial_\mu\,;
%\nonumber \\
& & g_{\mu\nu}(x) \rightarrow \hat g_{\mu\nu}(x) = e^{2\alpha(x)}g_{\mu\nu}(x)\,;\quad
\sqrt{-g(x)} \rightarrow \sqrt{-\hat g(x)} = e^{n\alpha(x)} \sqrt{-g(x)}\,;\nonumber\\
&& D_\mu \rightarrow \hat D_\mu\,;\quad  \Gamma^\lambda_{\mu\nu}(x) \rightarrow
\hat\Gamma^\lambda_{\mu\nu}(x)\,; \quad R_{\mu\nu}(x) \rightarrow \hat R_{\mu\nu}(x);
\quad R(x) \rightarrow \hat R(x)\,. \nonumber
\end{eqnarray}
With these replacements, the action--integrals of matter, vacuum and geometry
$A^M$, $A^G$ and $A^V$ described by Eqs (\ref{AM00})--(\ref{AG00}) are replaced by:
\begin{eqnarray}
\label{tildeAM00}
\hat A^M & = & \int\sqrt{-\hat g}\,\hat L^M\bigl(\hat g^{\mu\nu}, \hat {\boldsymbol\Psi},
\partial_\lambda \hat {\boldsymbol\Psi}\bigr)\,d^n x \,;\\
\label{tildeAG00}
\hat A^G  & = & -\frac{1}{2\kappa} \int\sqrt{-\hat g}\,\hat R\bigl(\hat g^{\mu\nu},
\partial_\lambda \hat g^{\mu\nu}\bigr)\, d^n x\,;\\
\label{tildeAV00}
\hat A^V & = & -\rho_{\hbox{\tiny vac}}\int\sqrt{-\hat g}\,d^n x \,.
\end{eqnarray}
It is thus evident that, when $e^{\alpha(x)}$ converges to 1 at the end of the inflation
epoch, all hat quantities of CGR converge to the homologous standard quantities and action integrals of GR.
Correspondingly, the EM tensors on the Cartan manifold
\begin{equation}
\label{tildeTheta_munu}
\hat \Theta^X_{\mu\nu}(x)=\frac{2}{\sqrt{-\hat g(x)}}\bigg[\frac{\delta \hat A^X}
{\delta \hat g^{\rho\sigma}(x)} - \partial_\lambda\frac{\delta \hat A^X}
{\delta\,\partial_\lambda \hat g^{\rho\sigma}(x)}\bigg]\,,
\vspace{3mm}
\end{equation}
where $X$ stands for $M$, $G$ or $V$, also converge to their homologous $\Theta^X_{\mu\nu}(x)$ of GR.
This makes sense because $\hat A^M$, $\hat A^G$ and $\hat A^V$ preserve the {\em formal properties}
of $A^M$, $A^G$ and $A^V$.

Since the way of passing from GR to CGR is so simple, the structure of CGR may seem rather simple, after all.
In fact, this is not so, because all complications related to scale--factor variability remain hidden. The main
advantage of this approach to CGR is that the expression of the inflated action integral preserves the same form
as that of GR.

\subsection{Conformal General Relativity on a Riemann manifold}
\label{CGROnRM}
In this subsection, we introduce an equivalent approach to CGR, which is more suitable for practical computations,
although less immediate as regard to physical interpretation. It consists of transferring the role played by the scale
factor $e^{\alpha(x)}$ of the fundamental tensor (\ref{fundtensor}) to a scalar field $\sigma(x)=\sigma_0 e^{\alpha(x)}$,
where $\sigma_0$ is a constant of dimension $-1$, to be interpreted as the promoter of the inflation process. Let us call it the
{\em dilation field}. The theory of CGR is thus grounded in a Riemann manifold of metric tensor
$g_{\mu\nu}(x)$, which imparts mathematical simplification and manifest conformal symmetry to the theory.

To realize how far--reaching this change of view is, let us consider what happens to the geometry
action integral $\hat A^G$ introduced in the previous subsection when the scale factor is regarded
as a field. Using Eq (\ref{RtotildeR2}) of the Appendix, we obtain
\vspace{-2mm}
$$
\sqrt{-\hat g}\,\hat R = \sqrt{-g}\,e^{n\alpha}\bigl\{ e^{-2\alpha}R -(n-1)\,e^{-4\alpha}\bigl[(n-6) (\partial^\rho
e^{\alpha})\,\partial_\rho e^{\alpha}+2\,D_\mu (e^{\alpha}\partial^\mu e^{\alpha})\bigr]\bigr\},
\vspace{-2mm}
$$
which converges to the Lagrangian density $\sqrt{-g(x)}\,R(x)$ of $A^G$ as $e^{\alpha(x)}$ converges to $1$.
In particular, for $n=4$, we have $\sqrt{-\hat g}\,\hat R = \sqrt{-g}\,\bigl[ e^{2\alpha}R +6\,(\partial^\rho
e^{\alpha})\,\partial_\rho e^{\alpha}-6\,D_\mu (e^{\alpha}\partial^\mu e^{\alpha})\bigr]$.
Therefore, by replacing $e^{\alpha(x)}$ with $\sigma(x)/\sigma_0$, Eq (\ref{tildeAG00}) takes the form
\vspace{-2mm}
$$
\hat A^G = -\frac{1}{2\kappa}\int\!\!\sqrt{-\hat g(x)}\,\hat R(x)\,d^4 x =  -\frac{6}{\kappa\,\sigma_0^2}
\int\!\!\sqrt{-g}\,\Bigl[\frac{1}{2}\,g^{\rho\tau}(\partial_\rho \sigma)(\partial_\tau \sigma)+\frac{\sigma^2 R}{12}\Bigr]
d^4 x\,,
$$
where we have removed the surface term $\sqrt{-g}\,D_\mu\bigl(\sigma \partial^\mu \sigma\bigr)\equiv
\partial_\mu\bigl(\sqrt{-g}\,\sigma \partial^\mu \sigma\bigr)$ by integration; this equivalence
being due to the covariant--divergence properties described in Eqs (\ref{covdiv}).

Thus, if we put $\sigma_0 =\sqrt{6/\kappa} \equiv \sqrt{6}\,M_{rP}\simeq 5.9654\times 10^{18}$ GeV,
$\tilde A^G$ takes the simple form
$$
A^G = -\int\!\!\sqrt{-g}\,\Bigl[\frac{1}{2}\,g^{\rho\tau}(\partial_\rho \sigma)\,
\partial_\tau \sigma+\frac{\sigma^2 R}{12}\Bigr] d^4 x\,,
$$
which represents the conformal--invariant Lagrangian density of a ghost scalar field $\sigma(x)$ interacting
with the gravitational field through the term $\sigma^2 R/12$. $\hat A^G$ is renamed
$A^G$, as the integration is carried out on a Riemann manifold. This result, {\em which is only possible
in four spacetime dimensions}, looks like a mathematical miracle, which confirms the importance of conformal symmetry
in the theory of gravitation.

Performing analogous substitutions in $\hat A^M$ and $\hat A^V$, we find that all
constants with dimension $k$ must be multiplied by $[\sigma(x)/\sigma_0]^k$. In particular:
all mass terms $m_k$ which may appear in the Lagrangian density are replaced by a $m_k\,\sigma(x)/\sigma_0$,
fields $\hat\Psi(x)$  of dimension $n$ are replaced by $\bigl[\sigma(x)/\sigma_0\bigr]^{n} \Psi(x)$
and $\rho_{\hbox{\tiny vac}}$ is replaced by $\rho_{\hbox{\tiny vac}}[\sigma(x)/\sigma_0]^n$.
It is then clear that $\hat A^V$ cannot be interpreted as the energy density of the vacuum but rather as a contribution
to the energy density of geometry.

Despite the presence of the gravitational coupling constant $\kappa\equiv M^{-2}_{rP}$ and other possible
dimensional constants, the total action integral $\hat A= \hat A^M+\hat A^G + \hat A^V$ grounded in a Cartan
manifold is {\em equivalent} to a total action integral $A$, grounded on a Riemann manifold, which is free
from dimensional parameters but in which a new field $\sigma(x)$ appears.

In \S\,\ref{confinvariance}, we prove that $A$ is manifestly invariant under the
{\em  group of local conformal transformations}. Because of this overall symmetry,
the study of the behavior of matter and geometry during the inflationary epoch is greatly facilitated.

The possible interaction of $\sigma(x)$ with all other fields $\boldsymbol{\Psi}(x)$ of the theory and $g_{\mu\nu}(x)$
may result in a total Lagrangian density, in which clear--cut separation into three components $A^M$, $A^G$, $A^V$ is
no longer possible.
%In particular, as already shown, $A^V$ cannot be regarded simply as the action integral of the physical vacuum.
For this reason, with all generality we may express the conformal--invariant action integral derived
from $\tilde A$ in the form
\begin{equation}
\label{AMVG}
A = \int \sqrt{-g(x)}\,L(x)\,d^4 x \equiv  \int \sqrt{-g}\,L\bigl(\sigma, \partial_\lambda\sigma,
\boldsymbol{\Psi}, \partial_\lambda \boldsymbol{\Psi}, g_{\mu\nu},
\partial_\lambda g_{\mu\nu}\bigr)\,d^4 x \,.
\end{equation}

In summary, the total action integral of CGR can be expressed in two different but equivalent ways:
either as a functional $\hat A$ of hat quantities grounded in a Cartan manifold of fundamental tensor
$\hat g_{\mu\nu}(x)$, and containing several dimensional constants, or as a functional $A$ of non--tilde
quantities and dilation field $\sigma(x)$ entirely free of dimensional constants and grounded in a
Riemann manifold of metric tensor $g_{\mu\nu}(x)$.

The relations between GR, CGR on Cartan manifold (CM) and CGR on Riemann manifold (RM)
condense into the diagram
\vspace{2mm}

\vbox to 50pt{
\centerline{
\begin{picture}(10,15)(0,5)
\put(-20,-20){\vector(1,1){23}}
\put(2.5,5){\small GR}
\put(-47.5,-8){\small $e^{\alpha(x)}\!\rightarrow \!1$}
\put(40,-20){\vector(-1,1){23}}
\put(10,-25){\vector(1,0){30}}
\put(31.5,-8){\small $\sigma(x)\!\!\rightarrow \!\sigma_0$}
\put(10,-25){\vector(-1,0){30}}
\put(-136,-28){\small CGR on Cartan manifold}
\put(-10,-22){\small equivalent}
\put(45,-28){\small CGR on Riemann manifold.}
\end{picture}
}}
\vspace{3mm}

The possibility of two equivalent representations mark a substantial difference between CGR and GR.
In \S\,4 of Part II, we discuss how and why these representations are related to
different interpretations of physical events with respect to inertial reference frames, themselves
undergoing the effects of the inflationary scale expansion and with respect to those of the
post--inflationary era, which behave as described in GR.

\section{Conformal--invariant action integrals}
\label{confinvariance}
In CGR on $n$D Riemann manifold, field dimensions are determined by the condition that the
dimension of conformal--invariant action integrand $\sqrt{-g(x)}\,L(x)$, where $L(x)$ is a Lagrangian
density, is zero. Accordingly, scalar fields have dimension $1-n/2$ and spinor fields
dimension $(1-n)/2$. Since spacetime parameters $x^\mu$ have dimension zero, partial
derivatives $\partial_\mu$ also have dimension zero. Note that, since $ds^2(x)= g_{\mu\nu}(x)
\,dx^\mu dx^\nu$ has dimension 2, $g_{\mu\nu}(x)$ has dimension 2, $\sqrt{-g(x)}$ dimension $n$
and Ricci scalar $R(x)$ dimension $-2$. For consistency with expressions such as
$\partial_\mu - igA_\mu(x)$, covariant gauge fields $A_\mu(x)$ have dimension zero. However,
since $g^{\mu\nu}(x)$ has dimension $-2$, by virtue of equation $g^{\mu\lambda}(x)\,g_{\lambda\nu}(x)
= \delta^\mu_\nu$, contravariant gauge fields $A^\mu(x)$ have dimension $-2$, etc.

To state the main properties of conformal invariant action integrals, the following little
theorem is of help:

A necessary (but not sufficient) condition for the action integral of a Lagrangian density $L(x)$, grounded
on a $n$D Riemann manifold of metric tensor $g^{\mu\nu}(x)$, to be conformal invariant is not only that it
is free of dimensional constants -- which is quite obvious since conformal invariance implies global
scale--invariance -- but also that the trace $\Theta(x)=g^{\mu\nu}(x)\,\Theta_{\mu\nu}(x)$ of
its EM tensor $\Theta_{\mu\nu}(x)$ vanishes.

{\em Proof:} Let $A=\int\sqrt{-g(x)}\,L(x)\,d^nx$ be the action integral. Conformal invariance
implies the vanishing of the variation of $A$ under infinitesimal Weyl transformations of the form
$$
g^{\mu\nu}(x)\rightarrow g^{\mu\nu}(x)+\delta_{\epsilon}\, g^{\mu\nu}(x) \equiv
g^{\mu\nu}(x)-2\,\epsilon(x)\, g^{\mu\nu}(x)\,,
$$
where $\epsilon (x)$ is an arbitrary infinitesimal function of  $x$.  We
therefore have
$$
\delta_\epsilon A = -2\int \sqrt{-g(x)}\,\epsilon(x)\,
g^{\mu\nu}(x)\,\Theta_{\mu\nu}(x)\,d^nx = -2\int \sqrt{-g(x)}\,\epsilon(x)\,
\Theta(x)\,d^nx =0\,,
$$
from which equation $\Theta(x)=0$ follows. %$\Box$

\subsection{Action integrals of gauge fields are conformal invariant only in 4D}
{\it Proof:} The Lagrangian density of a covariant Yang--Mills field $A^{Y\!M}_\mu$ in $n$D is
$$
L^{Y\!M}=-\frac{1}{4} \hbox{Tr}[{\boldsymbol F}_{\mu\nu}{\boldsymbol F}^{\mu\nu}]\,,\,\,
\hbox{with }{\boldsymbol F}_{\mu\nu}= {\boldsymbol \tau}_a F^a_{\mu\nu}\,,\,\, F^a_{\mu\nu}=
\partial_\mu A^a_\nu - \partial_\nu A^a_\mu + gf^a_{bc}A^b_\mu A^c_\mu\,,
$$
where ${\boldsymbol \tau}_a$ are group--generator matrices, $f^a_{bc}$ group
structure constants and $g$ the interaction constant, implying that the action
integral has dimension $n-4$. The symmetric EM tensor and its
trace are respectively
$$
\Theta^{Y\!M}_{\mu\nu} = -\frac{1}{4} \hbox{Tr}[{\boldsymbol F}_{\mu\lambda}{\boldsymbol F}^\lambda_\nu]
+\frac{1}{4}g_{\mu\nu}\hbox{Tr}[{\boldsymbol F}_{\mu\nu}{\boldsymbol F}^{\mu\nu}]\,,\quad
\Theta^{Y\!M} = \Bigl(\frac{n}{4}-1\Bigr)\hbox{Tr}[{\boldsymbol F}_{\mu\nu}\,{\boldsymbol F}^{\mu\nu}]\,.
$$
It is thus evident that conformal invariance holds only for $n=4$.

\subsection{Action integrals of massless spinor fields are conformal invariant in any dimension}
{\it Proof:} The Lagrangian density of a free fermion field $\psi$ of mass $M$ on the Riemann manifold has the form
\vspace{-2mm}
$$
L^F=\frac{i}{2}\Bigl[\bar\psi
(\slashed{D}\psi)-\overline{(\slashed{D}\psi)}\psi \Bigr] +M\bar\psi\psi\,,\,\,
\hbox{with } \slashed{D} \psi(x) = \gamma^\mu(x) \bigl[\partial_\mu
+\Gamma_{\mu}(x)\bigr]\,\psi(x)\,,
\vspace{-2mm}
$$
with $\gamma^\mu(x) = \gamma^a e^\mu_a(x)$, where $\gamma^a$ are the Dirac matrices in $n$D and $e_a^\mu(x)$ the
``soldering forms'', i.e., the $n$D analogs of tetrads in 4D. $\Gamma_\mu(x)$ are the (anti--Hermitian) spin matrices,
which are necessary to make partial derivatives $\partial_\mu$ covariant \cite{WEINBERG} \cite{WELDON}, and
$\overline{(\slashed{D}\,\psi)}$ and  $\bar\psi$ are the covariant Hermitian--conjugate of $\slashed{D}\,\psi$
and $\psi$, respectively. The motion equation is therefore $i\slashed{D}\, \psi(x)= M \psi(x)$ and the EM tensor is
$$
\Theta^F_{\mu\nu} = \frac{i}{4} \bigl[\bar \psi\gamma_\mu \slashed{D}_\nu\psi+
\bar \psi\gamma_\nu \slashed{D}_\mu\,\psi - (\overline{\slashed{D}_\mu\psi})
\gamma_\nu \psi + (\overline{\slashed{D}_\nu\psi}) \gamma_\mu \psi\bigr]\,.
$$
By index contraction of $\Theta^F_{\mu\nu}$, and exploitation of motion equations, we immediately obtain the trace
$\Theta^F=M\bar\psi\psi$, clearly implying conformal invariance for $M$ = 0 for any $n$.

Since for a gauge--field multiplet $A^a_\mu(x)$ and a spinor--field multiplet $\psi(x)$,
the expression $A^a_\mu(x)\bar\psi(x){\boldsymbol \tau}_a\gamma^\mu(x)\psi(x)$ is conformal
invariant, it is easy to prove that conformal--invariant Lagrangian densities of spinor
fields interacting with gauge fields exist only in 4D.

\subsection{Action integrals of scalar fields in a curved spacetime are conformal invariant only in 4D}
\label{Actintforvarphi}
{\it Proof:} In $n$D, the more general Lagrangian density of a scalar field $\varphi$, with self--interaction
constant $c$, interacting with the gravitational field through the metric tensor $g_{\mu\nu}(x)$ and its
derivatives, and free from dimensional constants, has the general form
$$
L^{(\varphi)} = \frac{1}{2}\Bigl[g^{\mu\nu}(\partial_\mu \varphi)\,\partial_\nu
\varphi +a\,\varphi^2R - \frac{c(n-2)}{n}\,\varphi^{2n/(n-2)}\Bigr]\,,
$$
where is $R$ the Ricci scalar and $a$ a suitable real constant.

The motion equation of $\varphi(x)$ is
\vspace{-2mm}
\begin{equation}
\label{nDphimoteq}
D^2\varphi-aR\varphi+ c\,\varphi^{(n+2)(n-2)}=0\,,
\vspace{-2mm}
\end{equation}
where $D^2=D^\mu D_\mu$ and $D_\mu$  are respectively the covariant d'Alembert operator and the covariant derivatives
constructed out of $g_{\mu\nu}$. The (improved) EM tensor \cite{CALLAN} is
\vspace{-1mm}
\begin{eqnarray}
\Theta^{(\varphi)}_{\mu\nu} &=& (\partial_\mu \varphi)\,\partial_\nu \varphi-
\frac{g_{\mu\nu}}{2}\Bigl[g^{\rho\sigma}(\partial_\rho
\varphi)\,\partial_\sigma \varphi-
\frac{c(n-2)}{n}\,\varphi^{2n/(n-2)}\Bigr]+\nonumber\\
\vspace{-2mm}
& & a\bigl(g_{\mu\nu}D^2-D_\mu\partial_\nu \bigr)\,\varphi^2 +
a\,\varphi^2\Bigl( R_{\mu\nu}-\frac{1}{2}\,g_{\mu\nu} R\Bigr) \,;\nonumber
\vspace{-2mm}
\end{eqnarray}
where Eq (\ref{Rvariation}) of Appendix is exploited, an integration by parts is performed, and a surface
term is suppressed.

By contraction with $g^{\mu\nu}$ and using motion equation (\ref{nDphimoteq}), we see that the trace of
the EM vanishes only if  $a=(n-2)/4(n-1)$, which implies that the action integral can be conformal invariant
only if the Lagrangian density has the general form
\vspace{-1mm}
$$
L^{(\varphi)} = \frac{1}{2}\Bigl[g^{\mu\nu}(\partial_\mu \varphi)\,\partial_\nu
\varphi +\frac{n-2}{4\,(n-1)}\,R\,\varphi^2 -
\frac{c\,(n-2)}{n}\,\varphi^{2n/(n-2)}\Bigr]\,.
\vspace{-1mm}
$$

Weyl transformations $\varphi(x)\rightarrow \bar \varphi(x) = e^{-\beta(x)}\varphi(x)$, $g^{\mu\nu}(x)
\rightarrow \bar g^{\mu\nu}(x) = e^{-2\beta(x)}g^{\mu\nu}(x)$, $\sqrt{-g(x)}  \rightarrow \sqrt{-\bar g(x)}
=  e^{n\beta(x)} \sqrt{-g(x)}$ and the first of (\ref{RtotildeR}), with $\bar R$ in place of $\hat R$ and
$\beta(x)$ in place of $\alpha(x)$, i.e., $\bar R  =  e^{-2\beta}R -(n-1)\,e^{-4\beta}\big[(n-4)
(\partial^\rho e^{\beta})\,\partial_\rho e^{\beta}+2\,e^{\beta} D^2 e^{\beta}\bigr]$, produce the transformation
$\sqrt{-g}\,L^{(\varphi)}\rightarrow \sqrt{-\bar g}\,\bar L^{(\varphi)} = \sqrt{-g}\,e^{(n-4)\beta}\big[L^{(\varphi)}
+ \Delta L^{(\varphi)}\big]$, where
\vspace{-1mm}
\begin{equation}
\label{DeltaL}
\Delta L^{(\varphi)}= \varphi^2(\partial_\mu\beta)\,\partial^\mu\beta -
(\partial_\mu \beta)\,\partial^\mu \varphi^2-\frac{n-2}{2}\,\varphi^2
e^{-\beta}D^2e^{\beta}- \frac{(n-4)(n-2)}{4}\varphi^2(\partial_\mu\beta)
\,\partial^\mu\beta.\nonumber
\end{equation}

Using the identity $D_\mu\big(\varphi^2 e^{-\beta}\partial^\mu e^{\beta}\big)=\varphi^2 e^{-\beta}D^2
e^{\beta} +(\partial_\mu\beta)\,\partial^\mu \varphi^2- \varphi^2 (\partial_\mu\beta)\,\partial^\mu\beta$ and the
covariant--divergence property $\sqrt{-g}\, D_\mu \big(\varphi^2 e^{-\beta}\partial^\mu e^{\beta}\big)
\equiv \partial_\mu\big(\sqrt{-g}\,\varphi^2 e^{-\beta}\partial^\mu e^{\beta}\big)$ stated in Eqs (\ref{covdiv}),
we immediately realize that, if $R\neq 0$, $\Delta L^{(\varphi)}$ is a surface term if and only if $n = 4$.

Instead, if $R = 0$, then it is $\beta = 0$ and consequently $\Delta L^{(\varphi)}= 0$. In this case, $L^{(\varphi)}$
is conformal invariant in any dimension.

In summary, the type of conformal--invariant Lagrangian density of a scalar field $\varphi$
in a curved spacetime is only possible in 4D and has the general form
\vspace{-1mm}
\begin{equation}
\label{Lvarphi}
L^{(\varphi)} = \frac{1}{2}\,g^{\mu\nu}(\partial_\mu \varphi)\,\partial_\nu \varphi+
\frac{R}{12}\,\varphi^2 - \frac{\lambda}{4}\,\varphi^4\,,
\vspace{-1mm}
\end{equation}
where $\lambda$ is a suitable constant. The motion equation of $\varphi(x)$ is then $D^2\varphi - R\,\varphi/6 +\lambda\,\varphi^3=0$.

Since for a scalar field $\varphi(x)$ and a spinor field $\psi(x)$, the expression $\sqrt{-g(x)}
\,\varphi(x)\bar\psi(x)\psi(x)$ is conformal invariant, it is easy to prove that conformal--invariant
Lagrangian densities of spinor fields interacting with scalar fields are only possible in 4D.

\subsection{The action integral of a ghost scalar field has a geometric meaning}
\label{scalarghosts}
Based on the results of the previous subsection, we might think that a gravitational equation similar
to that of GR comes spontaneously into play, provided that $\varphi$ has a non--zero vacuum expectation
value (VEV). Unfortunately, in this way the gravitational coupling constant would have the wrong sign, i.e., the
gravitational field would be repulsive.

Instead, let us assume that the action integral is negative, i.e., it has the form
\begin{equation}
\label{A^sigma}
A^{(\sigma)}=-\int \frac{\sqrt{-g}}{2}\Bigl[g^{\mu\nu}(\partial_\mu \sigma)\,\partial_\nu
\sigma+ R\frac{\sigma^2}{6}+ \frac{\bar \lambda}{2}\,\sigma^4\Bigr]d^4x\,,
\end{equation}
with $\bar \lambda$ a real constant and $\sigma(x)$ always positive (this is admissible
because the motion equation is invariant under $\sigma\rightarrow -\sigma$).
Therefore, without loss of generality, we can put $\sigma(x) = \sigma_0 \,e^{\alpha(x)}$
with $\sigma_0>0$. Note that, if $\bar\lambda>0$, the potential--energy density $\bar\lambda\,\sigma^4(x)/4$
is always positive, which may play an important role in moderating the infinite growth of
$\sigma(x)$  in suitable dynamic conditions. Since the kinetic energy of $A^{(\sigma)}$ is negative,
$\sigma(x)$ carries negative kinetic energy. Therefore it cannot be regarded as a physical field,
but rather as {\em a ghost scalar field potentially invested with geometric meaning}.
Since it causes the violent expansion of the geometric scale, we call it the {\em dilation field}.

{\em Note}: It is generally believed that the introduction of a scalar ghost in a quantum field
theory should cause the violation of $S$--matrix unitarity. In fact, this does not occur if
the ghost interacts with a physical scalar field so as to make the spectrum of the Hamiltonian
bounded from below \cite{ILHAN}. An interaction of this sort will be introduced in \S\,\ref{G2MneedsH}.

The Ricci--scalar factor of Eq.(\ref{A^sigma}) now has the right sign for gravity to be attractive.
Even better, let us put $\sigma_0 = \sqrt{6/\kappa}= \sqrt{6}\, M_{rP} \simeq 5.9654\times 10^{18}\,\hbox{GeV}$,
where $\kappa$ is the gravitational coupling constant and perform the following Weyl transformations:
$g^{\mu\nu}(x)\rightarrow \hat g^{\mu\nu}(x)= e^{-2\alpha(x)}g^{\mu\nu}(x)$, $\sqrt{-g(x)}\rightarrow
\sqrt{-\hat g(x)}= e^{4\alpha(x)}\sqrt{-g(x)}$, for any local quantity $Q_n(x)$ of dimension $n$,
$Q_n(x) \rightarrow \hat Q_n(x)= e^{n\alpha(x)}Q_n(x)$ and, in particular, $\sigma(x)\rightarrow
\hat \sigma(x)= e^{-\alpha(x)}\sigma(x) \equiv \sigma_0$. With these replacements $A^{(\sigma)}$ is
transformed into
\begin{equation}
\label{A^sigma0}
\hat A^{(\sigma_0)}=  - \int\frac{\sqrt{-\hat g(x)}}{2\kappa} \hat R(x)\,d^4x
- \frac{\bar\lambda\,\sigma_0^4}{4} \int\!\sqrt{-\hat g(x)}\,d^4x\,.
\end{equation}
which is {\em formally}, but not substantially, equal to the corresponding expression for the Einstein--Hilbert
action--integral of standard GR. Then, to the limit $e^{\alpha(x)}\rightarrow 1$, the term proportional
to $\bar\lambda\,\sigma_0^4/4$, if it is not canceled by other terms, gives an improper contribution to
the cosmological constant.

Thus, the conformal--invariant action integral $A^{(\sigma)}$ grounded in the Riemann manifold takes the form
of a non--conformal--invariant action integral $\hat A^{(\sigma_0)}$, for gravity and vacuum, grounded on
the Cartan manifold. Note that choosing the positive sign for $\sigma_0$ is formally equivalent to assuming
the spontaneous breakdown of conformal symmetry, in such a way that the degree of freedom of the ghost scalar
field is incorporated into the determinant of the fundamental tensor on the Cartan manifold.
We therefore have the following remarkable result:

For $n=4$, and only for $n=4$, the conformal--invariant action integral of a positive scalar--ghost
field $\sigma(x)= \sigma_0 e^{\alpha(x)}$ on a Riemann manifold of metric $g_{\mu\nu}(x)$ is equivalent
to a non--conformal--invariant action integral on a Cartan manifold with fundamental tensor
$\hat g_{\mu\nu}(x) = e^{2\alpha(x)}g_{\mu\nu}(x)$, in which the dimensional constant $\sigma_0$ plays
the role of conformal--symmetry--breaking parameter.

Unfortunately, Eq.(\ref{A^sigma}) makes sense only if it is part of a more complex action integral, as
the solution for $\sigma(x)$ to the motion equation derived from $A^{(\sigma)}$ is divergent. This raises the
question of how the expression of $A^{(\sigma)}$ could be appropriately included in a more general action integral
in order for the equation for  $\sigma(x)$ to be convergent. More general sorts of conformal--invariant geometry
action--integrals on 4D Riemann manifold may include a negative term proportional to the squared Weyl tensor $C^2(x)$,
of the form described by Eq.(\ref{weylterm}) of the Appendix, and a conformal--invariant potential--energy density term
$V(\sigma^2, {\boldsymbol\varphi})$, where ${\boldsymbol\varphi}$ is a suitable subset of all physical fields
accounting for any interactions between geometry and matter. That is, we assume that $V(\sigma^2, {\boldsymbol\varphi})$
should vanish if $\sigma$ or ${\boldsymbol\varphi}$ vanish. In view of this, it is suitable to represent the totality
of all matter fields by $({\boldsymbol \varphi}, {\boldsymbol\Psi})$, where ${\boldsymbol\Psi}$ stands for the subset
of matter fields different from ${\boldsymbol \varphi}$.

$V(\sigma^2, {\boldsymbol\varphi})$ is of degree two in $\sigma$ since, otherwise, motion equations may
permit $\sigma$ to change sign in the course of time, which would make it impossible to assume
$\sigma(x) =\sigma_0\,e^{\alpha(x)}$. Therefore, conformal--invariant interactions of fermions with
$\sigma$ must be excluded because such interactions are linear in $\sigma$.

Quadratic couplings with zero--mass vector fields as, for instance, $\sigma^2 g^{\mu\nu}A^a_\mu A^b_\nu$,  must
also be excluded, because a gauge--vector field cannot incorporate a scalar ghost as longitudinal spin component
via the Englert--Brout--Higgs mechanism \cite{ENGLERT0} \cite{HIGGS}. Thus, $V(\sigma^2, {\boldsymbol\varphi})$
can only depend on a set of real and/or complex massless scalar fields ${\boldsymbol \varphi}=\{\varphi_1,
\varphi_2,\dots \varphi_N\}$. Since all $\varphi_i$ have dimension one, the only possibility is to put $V(\sigma^2,
{\boldsymbol\varphi}) = \frac{1}{2} \sigma^2 c^{i j} \varphi^*_i \varphi_j$, where $[c^{i j}]$ is a suitable
$N\times N$ matrix (summation over repeated indices being implied).

In summary, the most general expression of a conformal--invariant geometry action--integral on the Riemann manifold
must have the form
\vspace{-1mm}
\begin{equation}
\label{A^GRU}  A^{(\sigma, {\boldsymbol \varphi})} = - \int \frac{\sqrt{-g}}{2}\bigg[\,g^{\mu\nu}(\partial_\mu \sigma)
\,\partial_\nu \sigma+\bigg(\frac{R}{6}- c^{i j} \varphi^*_i
\varphi_j\bigg)\,\sigma^2 + \frac{\bar \lambda}{2}\,\sigma^4\bigg]d^4x\,,
\vspace{-1mm}
\end{equation}
where $\varphi_i$ are massless scalar fields. Here, the Lagrangian--density term proportional to
the conformal--curvature tensor $C^2(x)$, as described by Eq.(\ref{weylterm}) of the Appendix, is
excluded for the reasons explained at the end of the Appendix.

Applying the Weyl transformations described in \S\,\ref{Cartanmanifolds} of the Appendix and already used in
Eq.(\ref{A^sigma}), we obtain the the most general expression of the geometry action--integral on the Cartan manifold
%\vspace{-1mm}
\begin{equation}
\label{ACG0} \hat A^{(\sigma_0, \hat {\boldsymbol \varphi})}  = - \int \sqrt{-\hat g}\,\bigg[\frac{1}{2\kappa}\,
\hat R - \frac{1}{2}\,\sigma^2_0\, c^{i j}\, \hat\varphi^*_i \hat\varphi_j + \frac{\bar \lambda}{2}\,\sigma^4_0\bigg]
d^4x\,.
\end{equation}

In this representation, the conformal symmetry of $A^{(\sigma, {\boldsymbol \varphi})}$ appears explicitly broken
by the dimensional constant $\sigma_0$. Since the explicit dependence of the action integral on the dilation field
has now disappeared, we may safely move the Lagrangian--density terms $\frac{1}{2}\sigma^2_0 c^{i j}
\hat\varphi^*_i \hat\varphi_j$ and $-\frac{\bar \lambda}{2}\,\sigma^4_0$ to the matter Lagrangian density.

As regards the action integral of matter on the Riemann manifold, let us denote by $A^{({\boldsymbol\varphi},
{\boldsymbol\Psi})}$ the part of total action integral which depends on all physical fields
$({\boldsymbol\varphi}, {\boldsymbol\Psi})$, i.e., not on $\sigma(x)$ via $V(\sigma^2, {\boldsymbol\varphi})$,
and by $L^{({\boldsymbol\varphi}, {\boldsymbol\Psi})}(x)$ its Lagrangian density. We
then have
%\vspace{-1mm}
\begin{equation}
\label{varphi&Psi}
A^{({\boldsymbol\varphi}, {\boldsymbol\Psi})} = \int\sqrt{-g(x)}\,L^{({\boldsymbol\varphi},
{\boldsymbol\Psi})}(x)\, d^4x\,,
%\vspace{-1mm}
\end{equation}
as well as the corresponding action integral on the Cartan manifold
%\vspace{-1mm}
\begin{equation}
\label{tildevarphi&Psi}
\hat A^{(\hat{\boldsymbol\varphi}, \hat{\boldsymbol\Psi})} = \int\sqrt{-\hat g(x)}
\,\hat L^{(\hat{\boldsymbol\varphi}, \hat{\boldsymbol\Psi})}(x)\, d^4x\,.
%\vspace{-1mm}
\end{equation}

The most general form of total action integral of matter and geometry on the Riemann and
Cartan manifolds can then be respectively written as
%\vspace{-1mm}
\begin{eqnarray}
\label{Atotal}
A & = & A^{({\boldsymbol\varphi}, {\boldsymbol\Psi})}+A^{(\sigma)} -\int \sqrt{-g(x)}\,
V\!\big[\sigma^2(x), {\boldsymbol\varphi}(x)\big]\, d^4x\,;\\
\label{tildeAtotal} \hat A & = & \hat A^{(\hat{\boldsymbol\varphi}, \hat{\boldsymbol\Psi})}
+\hat A^{(\sigma_0)}- \int\sqrt{-\hat g(x)}\,V\!\big[\sigma^2_0, \hat{\boldsymbol\varphi}(x)\big]\,d^4x\,;
%\vspace{-1mm}
\end{eqnarray}
with $A^{(\sigma)}$ and $\hat A^{(\sigma_0)}$ defined by Eqs. (\ref{A^sigma}) and  (\ref{A^sigma0}).

\newpage

Let us summarize here the most important aspects of the results so far achieved:
\begin{itemize}
\item[--]  Conformal invariance and the $4$-dimensionality of spacetime are closely related, since non--trivial
semiclassical conformal--invariant action--integrals in a curved spacetime  exist only in $4$D.
\item[-] Einstein's GR can be incorporated into CGR, provided that
conformal symmetry is spontaneously broken.

\item[--]  Matter-field Lagrangian densities on the Riemann and Cartan manifolds maintain the same algebraic form,
all quantities being replaced by hat quantities, whereas the form of the geometric Lagrangian density changes
considerably.

\item[--] In $4$D, and only in $4$D, the conformal symmetry of a conformal--invariant action integral on a Riemann
manifold containing a scalar ghost field $\sigma(x)= \sigma_0e^{\alpha(x)}$, $\sigma_0>0$, breaks down spontaneously
to an Einstein--Hilbert action integral on the Cartan manifold, with $g(x)= -[\sigma(x)/\sigma_0]^8$ playing the
role of the determinant of the fundamental tensor. Dimensional constant $\sigma_0$, which works as the conformal
symmetry-breaking parameter, is related to gravitational constant $\kappa$ by equation $\sigma^2_0 = 6/\kappa = 6\,M_{rP}^2$.
\end{itemize}

\section{Continuity equation for energy--momentum tensors}
\label{EMTconservation}
In this subsection, we prove that the invariance under diffeomorphisms of the GR action integral $A$, described in
\S\,\ref{Riemannmanifolds}, not only entails the gravitational equation, as already proven, but also the continuity
equation for the total EM tensor stated by covariant divergence equation $D^\mu\Theta_{\mu\nu}(x)=0$.

If $A$ depends not only on the set of internal fields  ${\boldsymbol\Phi(x)}$ but also on a set of external fields
${\boldsymbol\Psi}(x)= \{\Psi_1(x), \Psi_2(x),\dots, \Psi_m(x)\}$, the action integral can be written as
\begin{equation}
\label{Ahat}
A =  \int\sqrt{-g}\,L\big(g^{\mu\nu}, \partial_\lambda g^{\mu\nu}, {\boldsymbol\Phi},
\partial_\lambda {\boldsymbol\Phi}\big)\,d^4x - \int\sqrt{-g}\,U\big({\boldsymbol\Phi},{\boldsymbol\Psi}\big)\,d^4x \,;
\end{equation}
where $L(x)\equiv L\big(g^{\mu\nu}, \partial_\lambda g^{\mu\nu}, {\boldsymbol\Phi}, \partial_\lambda {\boldsymbol\Phi}\big)$
is the Lagrangian density described in \S\,\ref{Riemannmanifolds} and $U(x)\equiv \Delta L\big({\boldsymbol\Phi},
{\boldsymbol\Psi}\big)$ is the potential--energy density accounting for ${\boldsymbol\Phi}$--${\boldsymbol\Psi}$ interaction.

We need to know what the form of the continuity equation is in this case.

Under the action of an infinitesimal diffeomorphism $\delta_\varepsilon x^\mu$ of manifold coordinates,
$g^{\mu\nu}(x)$ and $\partial_\lambda g^{\mu\nu}(x)$ undergo the variation stated by Eq.(\ref{deltagupmunu})
and $\Psi_i(x)$ the variations
\vspace{-2mm}
\begin{equation}
\label{deltavarepPsi}
\delta_\varepsilon \Psi_j(x) = \varepsilon^\mu(x)\,\partial_\mu \Psi_j(x)\,.
\vspace{-2mm}
\end{equation}
We can therefore express the invariance of $A$ under arbitrary diffeomorphisms by
\vspace{-2mm}
\begin{equation}
\label{deltaepsA0}
\delta_\varepsilon A =\int \frac{\delta A}{\delta g^{\mu\nu}(x)}\,{\delta_\varepsilon g^{\mu\nu}(x)}\,d^4x
+\int \frac{\delta A}{\delta \Psi_j(x)}\,{\delta_\varepsilon \Psi_j(x)}\,d^4x =0.
\vspace{-2mm}
\end{equation}
Performing the variations of the single terms and removing the surface terms, we obtain
\vspace{-1mm}
\begin{eqnarray}
\label{deltaAM}
\hspace{-8mm}
\frac{\delta_\varepsilon  A}{\delta_\varepsilon g^{\mu\nu}(x)}  \!\!\!& = &  \!\!\!
\frac{\delta \bigl[\!\sqrt{-g(x)}\, L(x)\!\bigr]}{\delta g^{\mu\nu}(x)} -\partial_\lambda
\frac{\delta \bigl[\!\sqrt{-g(x)}\, L(x)\!\bigr]}{\delta[\partial_\lambda g^{\mu\nu}(x)]}\equiv
\frac{\sqrt{-g(x)}}{2}\,\Theta_{\mu\nu}(x);\\
\vspace{-2mm}
\label{deltaPsii}
\frac{\delta_\varepsilon A}{\delta_\varepsilon \Psi_j(x)} & = & -\sqrt{-g(x)}\,{\cal F}^j(x);
\vspace{-2mm}
\end{eqnarray}
where $\Theta_{\mu\nu}(x)$ is the EM tensor of $L(x)$ (according to Hilbert's definition) and ${\cal F}^i(x)$ are
the local forces generated by the spacetime variations of external fields.

The principle of GR thus yields
\vspace{-2mm}
\begin{equation}
\label{deltaepsA} \delta_\varepsilon A = \int\sqrt{-g(x)}\bigg\{\frac{1}{2}\,\Theta_{\mu\nu}(x)\,
\delta_\varepsilon g^{\mu\nu}(x)-\big[{\cal F}^i(x)\,\partial_\mu \Psi_j(x)\big]\varepsilon^\mu(x)\bigg\}\, d^4 x =0\,.
\vspace{-2mm}
\end{equation}
Now note that, inserting the second of Eqs (\ref{COVDER}) into Eq (\ref{deltagupmunu}), we obtain
\vspace{-2mm}
\begin{equation}
\label{deltavarepsgmunu}
\!\delta_\varepsilon g^{\mu\nu}(x) = \bigl[g^{\mu\rho}(x)\Gamma^\nu_{\lambda\rho}(x)+g^{\nu\rho}(x)
\Gamma^\mu_{\lambda\rho}(x)\bigr] \varepsilon^\lambda(x)+g^{\mu\rho}(x)\partial_\rho \varepsilon^\nu(x) +
g^{\nu\rho}(x)\partial_\rho \varepsilon^\mu(x)\,,
\vspace{-2mm}
\end{equation}
so that, exploiting the index symmetry of $\Theta_{\mu\nu}(x)$, we find the following equations
\vspace{-2mm}
\begin{eqnarray}
& &\!\!\!\!\frac{1}{2}\sqrt{-g(x)}\,\Theta_{\mu\nu}(x)\,\delta_\varepsilon g^{\mu\nu}(x)=
\sqrt{-g(x)}\,\Theta^\rho_{\,\cdot\,\nu}(x)\bigl[\,\partial_\rho \varepsilon^\nu(x)+\Gamma^\sigma_{\lambda\rho}(x)\,
\varepsilon^\lambda(x)\bigr]= \nonumber\\
\vspace{-2mm}
& &\!\!\!\!\partial_\rho\bigl[\sqrt{-g(x)}\,\Theta^{\rho}_{\,\cdot\,\nu}(x)\,\varepsilon^\nu(x)\bigr]-
D_\rho\bigl[\sqrt{-g(x)}\,\Theta^\rho_{\,\cdot\,\nu}(x)\bigr]\varepsilon^\nu(x)\,, \nonumber
\vspace{-2mm}
\end{eqnarray}
where $D_\rho\bigl[\sqrt{-g(x)}\,\Theta^\rho_{\,\cdot\,\nu}(x)\bigr]\equiv \partial_\rho \bigl[\sqrt{-g(x)}\,\Theta^{\rho}_{\cdot\,\nu}(x)\bigr] -
\sqrt{-g(x)}\,\Gamma^\sigma_{\rho\nu}(x)\Theta^{\rho}_{\cdot\, \sigma}(x)$ is used in the second step. Inserting this
into Eq (\ref{deltaepsA}), and removing the surface term, we obtain
\vspace{-2mm}
\begin{equation}
\delta_\varepsilon A =  -\int\sqrt{-g(x)}\Big[ D^\mu\,\Theta_{\mu\nu}(x)-{\cal F}^i(x)\,\partial_\mu \Psi_i(x)
\Big]\,\varepsilon^\nu(x)\,d^4 x=0\,,\nonumber
\vspace{-2mm}
\end{equation}
from which we can extract the generalized continuity equation
\vspace{-2mm}
\begin{equation}
\label{DmuThetamunu}
D^\mu\Theta_{\mu\nu}(x)={\cal F}^i(x)\,\partial_\nu \Psi_i(x)\,.
\vspace{-2mm}
\end{equation}
Clearly, the left side of this equation represents the power delivered by the external fields in the form of a source term for
EM--current density $\Theta_{\mu\nu}(x)$.

These results can be extended to CGR, in which case $\hat\Theta_{\mu\nu}(x)$ is replaced by the conformal EM tensor
defined by Eq.(\ref{tildeTheta_munu}), $D^\mu$ by $\hat D^\mu$, as shown by Eqs.(\ref{tildeDmutildeTheta}) and (\ref{tildeDvT2})
of the Appendix, Eq.(\ref{DmuThetamunu}) by $\hat D^\mu\hat\Theta_{\mu\nu}(x)={\cal \hat F}^i(x)\,\partial_\mu \hat\Psi_i(x)$,
and Eq.(\ref{deltavarepsgmunu}) by
\vspace{-2mm}
\begin{equation}
\label{tildedeltavarepsgmunu}
\delta_\varepsilon \hat g^{\mu\nu}(x) = \big[\,\hat g^{\mu\rho}(x)\hat \Gamma^\nu_{\lambda\rho}(x)+\hat g^{\nu\rho}(x)
\hat\Gamma^\mu_{\lambda\rho}(x)\big] \varepsilon^\lambda(x)+\hat g^{\mu\rho}(x)\partial_\rho \varepsilon^\nu(x) +
\hat g^{\nu\rho}(x)\partial_\rho \varepsilon^\mu(x).\nonumber
\vspace{-2mm}
\end{equation}

\subsection{Energy--momentum transportation}
\label{EMpartition}
Let $A^A_0$ and $A^B_0$ be the action integrals of two non--interacting scalar fields $\varphi^A(x)$ and
$\varphi^B(x)$ grounded in the same Riemann manifold, $L^A_0(x)$ and $L^B_0(x)$ their respective Lagrangian
densities, $\Theta^A_{0\,\mu\nu}(x)$ and $\Theta^B_{0\,\mu\nu}(x)$ their respective EM tensors.
As proven in the previous section, invariance of $A^A_0$ and $A^B_0$ under manifold diffeomorphisms
entails continuity equations $D^\mu\Theta^A_{0\,\mu\nu}(x)=0$ and $D^\mu\Theta^B_{0\,\mu\nu}(x)=0$.
We need to study how these equations change when the two fields interact through a potential--energy density
term $U(x)\equiv U\bigl[\varphi^A(x), \varphi^B(x)\bigr]$ obeying conditions
$U\bigl[\varphi^A(x), 0 \bigr]= U\bigl[0, \varphi^B(x)\bigr]=0$. In this case, the total action integral,
Lagrangian density and EM tensor become respectively: $A = A^A_0+A^B_0 -\int \sqrt{-g(x)}\,U(x)\, d^4x$,
$L(x) = L^A_0(x)+L^B_0(x) - U(x)$ and $\Theta_{\mu\nu}(x) = \Theta^A_{0\,\mu\nu}(x)+\Theta^B_{0\,\mu\nu}(x) +
g_{\mu\nu}(x)\,U(x)$, and the continuity equation takes the form
\begin{equation}
\label{DmuTheta}
D^\mu\Theta_{\mu\nu}(x)= D^\mu\Theta^A_{0\,\mu\nu}(x) + D^\mu\Theta^B_{0\,\mu\nu}(x) +
\partial_\nu U(x)=0\,.
\end{equation}
Therefore, using equation
\begin{equation}
\partial_\nu U(x) = \frac{\partial U(x)}{\partial \varphi^A(x)}\,\partial_\nu\varphi^A(x) +
\frac{\partial U(x)}{\partial \varphi^B(x)}\,\partial_\nu\varphi^B(x)\,,\nonumber
\end{equation}
we can split Eq (\ref{DmuTheta}) as follows
\begin{equation}
\label{DmuThetaA&B}
\Big[D^\mu\Theta^A_{0\,\mu\nu}(x) + \frac{\partial U(x)}{\partial \varphi^B(x)}\,\partial_\nu \varphi^B(x)\Big]+
\Big[D^\mu\Theta^B_{0\,\mu\nu}(x) + \frac{\partial U(x)}{\partial \varphi^A(x)}\,\partial_\nu \varphi^A(x)\Big]=0\,.
\end{equation}

If we now regard $\varphi^B(x)$ and $\varphi^A(x)$ respectively as the external fields of action integrals
$A^A = A^A_0-\int \sqrt{-g(x)}\,U(x)\, d^4x$ and $ A^B = A^B_0-\int \sqrt{-g(x)}\,U(x)\, d^4x$, separately
considered, then, based on the arguments discussed in the previous section and, in particular, on Eq.(\ref{DmuThetamunu}),
we can identify ${\cal F}^B(x)$ with $-\delta U(x)/\delta \varphi^B(x)$ and ${\cal F}^A(x)$
with $-\delta U(x)/\delta \varphi^A(x)$, separate the terms in squared brackets
of Eq (\ref{DmuThetaA&B}) and write the non--conservative continuity equations as
\begin{equation}
\label{divThetaABnumu}
D^\mu\Theta^A_{\mu\nu}(x) =  {\cal F}^B(x)\,\partial_\nu\varphi^B(x)\,;\quad
D^\mu\Theta^B_{\mu\nu}(x) =  {\cal F}^A(x)\,\partial_\nu\varphi^A(x)\,.
\end{equation}
The second members of this equation can be interpreted as the amount of EM-tensor charge
locally exchanged between $\Theta^B_{\mu\nu}(x)$ and $\Theta^A_{\mu\nu}(x)$ during the
$\varphi^A(x)$--$\varphi^B(x)$ interaction.

\section{Geometry--to--matter energy--transfer needs a Higgs field}
\label{G2MneedsH}
According to the results of the previous subsection, we can now easily determine the mechanism of energy transfer
from geometry to matter. Let us reconsider the complete action integral of matter and geometry introduced in Eq (\ref{Atotal}),
i.e.,
\vspace{-2mm}
$$
A = A^{({\boldsymbol\varphi},{\boldsymbol\Psi})}+A^{(\sigma)} -\int\sqrt{-g(x)}\,V(x)\,d^4x\,,
\vspace{-2mm}
$$
where $A^{({\boldsymbol\varphi}, {\boldsymbol\Psi})}$ is the action integral of matter described in Eq (\ref{varphi&Psi}),
$A^{(\sigma)}$ is the action  integral of geometry described in Eq (\ref{A^sigma}) and $V(x) = \frac{1}{2}\,g_{\mu\nu}(x)
\,\sigma^2(x)\, c^{i j}\varphi^*_i(x)\,\varphi_j(x)$ is the potential--energy density accounting for the interaction of
ghost scalar field $\sigma(x)$  with a $n$ real and/or complex scalar fields $\{\varphi_1(x), \dots, \varphi_n(x)\}$.
The EM tensor of $A$ is thus:
\vspace{-1mm}
$$
\Theta_{\mu\nu}(x) = \Theta^{({\boldsymbol\varphi}, {\boldsymbol\Psi})}_{\mu\nu}(x) + \Theta^{(\sigma)}_{\mu\nu}(x) +\frac{1}{2}\,g_{\mu\nu}(x)\,\sigma^2(x)\, c^{i j}\varphi^*_i(x)\,\varphi_j(x)\,,
\vspace{-1mm}
$$
where $\Theta^{({\boldsymbol\varphi}, {\boldsymbol\Psi})}_{\mu\nu}(x)$ and $\Theta^{(\sigma)}_{\mu\nu}(x)$
are the contributions to $\Theta_{\mu\nu}(x)$ respectively from action integrals $A^{({\boldsymbol\varphi},
{\boldsymbol\Psi})}$ and $A^{(\sigma)}$ and $[c^{i j}]$ is a hermitian matrix. In accordance with Eq (\ref{DmuTheta}),
the EM conservation equation has the form
\vspace{-2mm}
\begin{equation}
\label{ThetamunuMatter&sigma}
D^\mu\Theta_{\mu\nu}(x)=  D^\mu\Theta^{({\boldsymbol\varphi},{\boldsymbol\Psi})}_{\mu\nu}(x) +
D^\mu\Theta^{(\sigma)}_{\mu\nu}(x) +\frac{1}{2}\,\partial_\nu\,\sigma^2(x)\, c^{i j}\varphi^*_i(x)\,\varphi_j(x) = 0\,.
\vspace{-1mm}
\end{equation}

As described in the previous subsection, we can extract from $A$ the partial action integrals
\vspace{-1mm}
\begin{equation}
\label{PartAMG}
A^M =  A^{({\boldsymbol\varphi}, {\boldsymbol\Psi})}- \int\sqrt{-g(x)}\,V(x)\,d^4x,\quad
A^G  =  A^{(\sigma)} - \int\sqrt{-g(x)}\,V(x)\,d^4x\,,
\vspace{-1mm}
\end{equation}
to be interpreted respectively as the action integral of matter interacting with $\sigma(x)$, now regarded
as an external field, and that of geometry interacting with $\{{\boldsymbol\varphi}, {\boldsymbol\varphi}^*\}$, now
regarded as external fields. In conformity with Eqs (\ref{divThetaABnumu}), we can then split
Eq (\ref{ThetamunuMatter&sigma}) as follows,
\vspace{-2mm}
\begin{equation}
\label{DmuThetaM&G}
D^\mu\Theta^M(x)= \frac{1}{2} \sigma^2(x)\, c^{ij}\partial_\nu \big[\varphi_j(x)\,\varphi^*_j(x)\big],\,\,\,
D^\mu\Theta^G_{\mu\nu}(x) =  \frac{1}{2}\,c^{ij}\,\varphi^*_i(x)\varphi_j(x)\,\partial_\nu \sigma^2(x),
\vspace{-2mm}
\end{equation}
respectively describing the rates of EM transfer from geometry to matter and vice versa, the first of which is always
positive, provided that all the eigenvalues of $[c^{ij}]$ are positive.

It is therefore evident that, in order for CGR to approach GR at the infinite time limit, the second members of
Eqs (\ref{DmuThetaM&G}) must approach zero  at the same limit, implying that both $\sigma(x)$ and $\varphi_i(x)$
tend to become constant in the course of time.

To clarify this point, let us focus on a conformal--invariant action integral in which only
$\{{\varphi}_i, {\varphi}^*_i\}$ and $\sigma$ exist, namely, an action integral of the general form
\vspace{-2mm}
\begin{eqnarray}
A & = & \int \frac{\sqrt{-g}}{2}\Bigl[g^{\mu\nu}\delta^{ij}(\partial_\mu\varphi^*_i)(\partial_\nu\varphi_j)
 - Q(\varphi, \varphi^*) +\frac{R}{6}\,\delta^{ij}\varphi^*_i\varphi_j
 + \nonumber\\
& & \sigma^2 c^{ij}\varphi^*_i\varphi_j- g^{\mu\nu}(\partial_\mu\sigma)(\partial_\nu\sigma)-
\frac{\bar\lambda}{2}\, \sigma^4 - \frac{R}{6}\,\sigma^2 \Bigr]d^4x\quad (i,j = 1,2 \dots N)\,,\nonumber
\vspace{-2mm}
\end{eqnarray}
where $\delta^{ij}$ is the Kronecker delta and $Q(\varphi, \varphi^*)$ is a real polynomial of
fourth degree in $\varphi_i$ and $\varphi^*_i$, which represents self--interactions and
possible mutual interactions of these fields.

Since many different choices are {\em a priori} possible for $c^{ij}$ and $Q(\varphi, \varphi^*)$, in the
absence of any sufficient reason in favor of a particular choice, we invoke the heuristic principle of
maximum symmetry of fundamental laws by assuming that all $\varphi_i$ are complex, $c^{ij}=
c\, \delta^{ij}$, $Q(\varphi, \varphi^*)= \frac{1}{2}\,\lambda\, |{\boldsymbol\varphi}|^4$,
where $|{\boldsymbol\varphi}|^2 = \delta^{ij}\varphi^*_i\varphi_j=\sum_i \vert\varphi_i\vert^2$, with $c$
and $\lambda$ suitable real constants. This makes it possible that $2N-1$ massless degrees of freedom
of vector ${\boldsymbol\varphi}$ play the role of Nambu--Goldstone bosons giving mass to gauge vector--fields.
Hence we have
\begin{eqnarray}
\label{maxsymmA}
A &\!\! = \!\! & \int \frac{\sqrt{-g}}{2}\Bigl[\textstyle\sum_i g^{\mu\nu}(\partial_\mu\varphi^*_i)(\partial_\nu\varphi_i) -\displaystyle \frac{\lambda}{2}|{\boldsymbol\varphi}|^4  + c\,\sigma^2 |{\boldsymbol\varphi}|^2 - \frac{\bar\lambda}{2}\, \sigma^4+\nonumber\\
& &
\frac{R}{6}(|{\boldsymbol\varphi}|^2-\sigma^2)-
g^{\mu\nu}(\partial_\mu\sigma)(\partial_\nu\sigma)\Bigr]d^4x\,,
\end{eqnarray}
from which we derive the motion equations
$$
D^2 \sigma - \frac{R}{6}\,\sigma -\bar \lambda\, \sigma^3 + c\, \sigma \,|{\boldsymbol\varphi}|^2 =0\,;\quad
D^2 \varphi_i - \frac{R}{6}\,\varphi_i +\lambda\,|{\boldsymbol\varphi}|^2 \varphi_i
- c\, \sigma^2\varphi_i=0 \,\,\, \hbox{and c.c.}
$$

Temporal convergence to constant values of $\sigma$ and $\varphi_i$ for arbitrary initial states is possible
provided that $R$ also converges to a constant and $c\,|{\boldsymbol\varphi}|^2 -\bar\lambda\,\sigma^2 =
\lambda|{\boldsymbol\varphi}|^2 -c\,\sigma^2  = R/6$.

Hence, either $c=\lambda=\bar\lambda$ and $R=0$, or $\bar \lambda = c^2/\lambda$ and $R\neq 0$.
However, in both cases, we find that the potential--energy density of  $\sigma$--${\boldsymbol\varphi}$
interaction has the general form
$$
U\big(\sigma, {\boldsymbol\varphi}\big) = \frac{\lambda}{4}\bigg(|{\boldsymbol\varphi}|^2 - \frac{c}{\lambda}\,\sigma^2\bigg)^2\,,
$$
which, as a function of ${\boldsymbol\varphi}(x)$, exhibits a Mexican--hat profile of depth $\sqrt{c/\lambda}\,\sigma(x)$.

In the absence of any interaction of $\varphi_i(x)$ with gauge vector--fields, by a suitable unitary transformation of
matrix $[c^{ij}\,]$, we can bring ${\boldsymbol\varphi}(x)$ to the standard form ${\boldsymbol\varphi}(x) =
\{0,\dots,0,\varphi(x)\}$, where $\varphi(x)= |{\boldsymbol\varphi}(x)|$. In this case, Eq.(\ref{maxsymmA}) can be
simply written as
\begin{eqnarray}
\label{HiggsactintOnR}
\!\!A =\! \int\! \frac{\sqrt{-g}}{2}\bigg[g^{\mu\nu}\!\big(\partial_\mu\varphi\big)\,\partial_\nu\varphi
-g^{\mu\nu}\!\big(\partial_\mu\sigma\big)\,\partial_\nu\sigma\! -\!
\frac{\lambda}{2}\bigg(\varphi^2 -\!\frac{c}{\lambda}\,\sigma^2\bigg)^2 \!+\!
\frac{R}{6}\,\big(\varphi^2\!-\!\sigma^2\big)\bigg]d^4x.
\end{eqnarray}

We can see that $\varphi(x)$ behaves like a Higgs field of mass proportional to $\sigma(x)$, while the $2N-1$ corollary
Nambu--Goldstone bosons of ${\boldsymbol\varphi}(x)$ are transferred, once and for all, to the gauge vector sector
via an interaction term proportional to $A^k_\mu(x)\,{\boldsymbol\tau}_k {\boldsymbol\varphi}(x)$, where $A^k_\mu(x)$ are $2N-1$ gauge
vector--fields and ${\boldsymbol\tau}_k$ are the Lie algebra generators of $SU(N)$, so as to promote the spontaneous
breakdown of the Standard Model symmetries \cite{LANGACKER}. In these conditions, $R(x)$ and $g_{\mu\nu}(x)$ generally
depend on spacetime curvature and gravitational field. If the matter field is homogeneous and isotropic, as is presumably
the case during inflation and at infinite time, $R$ is constant and the gravitational field vanishes.

Carrying out the replacements
\vspace{-2mm}
\begin{eqnarray}
\label{Riemann2Cartan}
&&\hspace{-5mm}\sqrt{-g}\rightarrow \sqrt{-\hat g}= e^{4\alpha}\sqrt{-g},\quad g^{\mu\nu}\rightarrow \hat g^{\mu\nu} =
e^{-2\alpha}\hat g^{\mu\nu}, \quad \varphi\rightarrow \hat \varphi =e^{-\alpha}\varphi,\nonumber \\
&&\hspace{-5mm}\sigma\rightarrow \hat\sigma=\sigma_0,\quad R\rightarrow \hat R =e^{-2\alpha}\big(\hat R -6\,\sigma^{-1} D^2\sigma\big)\equiv
e^{-2\alpha}\big(\hat R- 6\,e^{-\alpha}D^2e^{\alpha}\big)
\vspace{-2mm}
\end{eqnarray}
in Eq (\ref{HiggsactintOnR}), which includes the Eq (\ref{TildeR4}) proven in the Appendix, we obtain the representation
of $A$ on the Cartan manifold:
\begin{equation}
\label{HiggsactintOnC}
\hat A =  \int \frac{\sqrt{-\hat g}}{2}\bigg[\hat g^{\mu\nu}\big(\partial_\mu\hat \varphi\big)\,
\partial_\nu\hat\varphi-\frac{\lambda}{2}\bigg(\hat \varphi^2 -\frac{c}{\lambda}\,\sigma_0^2\bigg)^2 -
\frac{\sigma_0^2}{6} \hat R \,\bigg(1- \frac{\hat\varphi^2}{\sigma_0^2}\bigg)\bigg]\,d^4 x.
\end{equation}
Let us show that action integrals (\ref{HiggsactintOnR}) and (\ref{HiggsactintOnC}) differ only by a surface term,
or, to say it differently, they are functionally equivalent. In effect, using the equalities of Eqs  (\ref{Riemann2Cartan}), we see
that Eq (\ref{HiggsactintOnC}) can be written as $\hat A = A +\Delta A$, where
\begin{equation}
\label{DeltaA}
\Delta A  = \int \frac{\sqrt{-g}}{2}\Big\{(\varphi^2-\sigma^2)\big[g^{\mu\nu}\big(\partial_\mu \alpha\big)\,\partial_\nu\alpha
-e^{-\alpha}D^2 e^{\alpha}\big]-g^{\mu\nu}\big(\partial_\mu \alpha\big)\,\partial_\nu\big(\varphi^2-\sigma^2\big)\Big\} d^4 x.
\end{equation}
Therefore, using identity
$$
D_\mu\big[(\varphi^2 - \sigma^2) e^{-\alpha}\partial^\mu e^\alpha\big]
= g^{\mu\nu}(\partial_\nu \alpha)\partial_\mu(\varphi^2 - \sigma^2) + (\varphi^2 - \sigma^2)\big[ e^{-\alpha}D^2 e^\alpha
- g^{\mu\nu}(\partial_\mu \alpha)\partial_\mu \alpha\big]
$$
and the covariant--divergence property $\sqrt{-g}\,D_\mu f^\mu  = \partial_\mu\big(\sqrt{-g}\,f^\mu\big)$, where $f^\mu$
is any contra\-variant vector [cf Eq (\ref{covdiv})], we can realize that $\Delta A$ is a mere surface term.

However, at variance with $A$, the conformal invariance of $\hat A$ is no longer manifest. In effect, passing
from $A$ to $\hat A$, the original conformal invariance of the former appears in the latter explicitly broken by constant
$\sigma_0= \sqrt{6}\,M_{rP}$, where $M_{rP} =2.4354\times 10^{18}$ GeV is the reduced Planck mass.

Note that, if we put $c=\mu^2_H/2\sigma_0^2$, where $\mu_H \simeq 126$ GeV the Higgs boson mass, $\hat A$ acquires
the formal properties of the action integral of a Higgs field of mass $\mu_H$ interacting with the gravitational field
through $\hat g^{\mu\nu}(x)$ and the term proportional to $\hat R(x)$, i.e.,
\begin{equation}
\label{asymptildeA}
\hat A =\int \frac{\sqrt{-\hat g}}{2}\bigg[\,\hat g^{\mu\nu}(\partial_\mu\hat\varphi)\,\partial_\nu\hat\varphi-
\frac{\lambda}{2}\Bigl(\hat\varphi^2 -\frac{\mu^2_H}{2\lambda}\Bigr)^2 -
\frac{\hat R}{\kappa}\Big(1- \frac{\hat\varphi^2}{\sigma_0^2}\Big) \bigg]d^4 x\,.
\end{equation}
Since we expect $\hat\varphi(x)$ to oscillate in the interval $[0,\hat\varphi_{\hbox{\tiny max}}]$, where
$\hat\varphi_{\hbox{\tiny max}} \leq  \mu_H/\sqrt{\lambda}\ll\sigma_0$, we see that
$\hat\varphi^2(x)/\sigma_0^2$ remains absolutely negligible relative to $1$, so that the gravitational interaction
term $-(1 -\hat \varphi^2/\sigma_0^2)\,\hat R/\kappa$ does not differ appreciably from $-\hat R/\kappa$.

Thus, if in the course of time $e^{\alpha(x)}$ approaches 1, then $\hat g^{\mu\nu}(x)$ approaches
$g^{\mu\nu}(x)$, $\hat\varphi(x)$ approaches $\varphi(x)$, $\hat R(x)$ approaches $R(x)$ and
therefore Eq (\ref{asymptildeA}) approaches the standard action integral of the Higgs field
$\varphi(x)$ interacting with the gravitational field, i.e.,
$$
A^{GR} =\int \frac{\sqrt{-g}}{2}\,\bigg[g^{\mu\nu}(\partial_\mu\varphi)\,\partial_\nu\varphi-
\frac{\lambda}{2}\Bigl(\varphi^2 -\frac{\mu^2_H}{2\lambda}\Bigr)^2 -
\frac{R}{\kappa} \bigg]d^4 x\,.
$$

In summary, the energy transfer from geometry to matter can only be explained
in the general framework of CGR, provided that the following two conditions are satisfied:
(1) the matter Lagrangian density of the action integral $A$ on the Riemann manifold
includes one or more massless scalar fields quadratically coupled with the dilation field;
(2) the corresponding Lagrangian density of the action integral $\hat A$ on the Cartan manifold
is formally equal to that of a Higgs boson field equipped with a number of corollary Goldstone
bosons, so that CGR is potentially capable of joining the Standard Model smoothly.

\section{The Mach principle according to G\"ursey}
\label{MEGursey}
As discussed in \S\,\ref{Cartanmanifolds}, the fundamental tensor of a 4D Cartan manifold can be written as
$\hat g_{\mu\nu}(x) = e^{2\alpha(x)}\, g_{\mu\nu}(x)$, with $\sqrt{-g(x)}=1$, where $\alpha(x)$ contains information
about the geometry of the universe on the large scale, whereas $g_{\mu\nu}(x)$ contains information about gravitational
interactions. The physical meaning of this factorization was clarified by G{\"u}rsey in 1963.

According to Mach--Einstein doctrine, here referred to as the {\em Mach Principle}, in the universe, there
exists a basic inertial frame which is globally determined by the distant bodies. Initially, this was
called the reference frame of ``fixed stars", but today it should be more properly called the reference frame
of galaxy clusters on the large scale. The existence of such a frame is ensured by the observed simplicity
of the universe on a sufficiently large scale. Unfortunately, this principle cannot be derived from
Einstein's gravitational equations since, in the theoretical framework of GR, the average effect of
distant bodies is unpredictable, because the structure of the universe on the large scale is {\em a priori}
undetermined.

\subsection{The conformal background of the universe}
\label{confbackgrnd}
To overcome the difficulty and implement the Mach principle within the framework of a new reformulation of GR,
G{\"u}rsey resorted to approaching the problem in three steps: (1) ``{\em to find a way of
separating local effects from the general cosmological structure due to the distribution of distant bodies,
because all statements related to Mach's principle involve such separation}; (2) {\em the boundary
conditions being only meaningful in a definite coordinate system, we must be able to introduce privileged
coordinate frames determined by the over--all cosmological structure that has been separated in the
first step. These are the inertial frames that, according to Mach, are determined, to within a
kinematical group, by the over--all distribution of matter}; (3) {\em to preserve the general
covariance, we have to show that Machian boundary conditions can also be generalized to an arbitrary
coordinate system, that is, to noninertial frames}''.

To satisfy these requirements, G{\"u}rsey hypothesized that the fundamental tensor of spacetime geometry
has a part $C_{\mu\nu}(x)= K(x)\,\gamma_{\mu\nu}(x)$, where $\gamma_{\mu\nu}(x)$ is the average metric
of the universe and $K(x)>0$ a homogeneous and isotropic scalar density \cite{WALD}, plus a second part describing
the deviations from this uniform structure.

The frame of distant bodies may therefor be defined as one in which $C_{\mu\nu}(x)$ is conformally flat,
which means that $\gamma_{\mu\nu}(x)$ is flat, so that light in this system
travels along a straight line of constant velocity $c \,(\, =1)$. The boundary conditions for the metric then
require that, with respect to the inertial frames of comoving observers, the fundamental tensor tends
asymptotically to a conformal metric. In this way, a transformation which takes the observer
from an inertial to a non--inertial frame may be interpreted as a transformation which distorts the
aspect of the cosmological background in such a way as to resemble effects due to accelerations with
respect to the ``fixed--star'' reference frame.

In G{\"u}rsey's view, it is not this relative acceleration which produces the Machian forces, but
is the way in which $K(x)$ depends on $x^\mu$ which defines the inertial behavior of the observer with
respect to the reference frame of heavy bodies on large scales, so that these bodies appear to be fixed when
$K(x)$ is homogeneous and isotropic.

The only fundamental tensor consistent with this view is $\hat g_{\mu\nu}(x) \equiv e^{2\alpha(x)}g_{\mu\nu}(x)$,
with the unimodularity condition $\sqrt{-g(x)}=1$. The factor $g_{\mu\nu}(x)$ can then be written as
$$
g_{\mu\nu}(x)= \gamma_{\mu\nu}(x)+h_{\mu\nu}(x)\,, \,\,\, \hbox{with\,\,
 det[$\gamma_{\mu\nu}(x)$] = det[$\gamma_{\mu\nu}(x)+h_{\mu\nu}(x)]$}=1\,,
$$
where $h_{\mu\nu}(x)$ describes the gravitational field as a deviation from $\gamma_{\mu\nu}(x)$. Thus, the
information on the over--all cosmological structure is contained in $C_{\mu\nu}(x)= e^{2\alpha(x)}\gamma_{\mu\nu}(x)$
whereas that on the gravitational field is contained  in $h_{\mu\nu}(x)$. For our needs we can assume $h_{\mu\nu}(x)$
to be very small, as if the gravitational field were a small linear perturbation of the background metric,
in which case the unimodularity condition implies the harmonic gauge property $\gamma^{\mu\nu}\,h_{\mu\nu}(x)=0$. This means, in practice,
that non--linear gravitational effects are ignored and possible black holes are replaced by extended bodies of large mass.

To fulfill the Mach principle, G{\"u}rsey assumed that $\hat g_{\mu\nu}(x)$ obeys
the following boundary condition at spatial infinity in the inertial frame
\begin{equation}
\label{gurseybound}
\hat g_{\mu\nu}(x) \rightarrow e^{2\alpha(\tau)}\,\gamma_{\mu\nu}(x)\,,
\end{equation}
where $\tau$ is the kinematic--time coordinate of the world lines stemming from the origin of a light--cone
to which the universe is confined, and
\begin{equation}
\label{dSfactor}
e^{\alpha(\tau)}= e^{\alpha(0)}\frac{1}{1+ R\,\tau^2/12}\,,
\end{equation}
where $R > 0$. This clearly implies that the spacetime is closed and has the structure of
deSitter space $dS_4$. Since this conflicts with the observed accelerated expansion of the universe, which implies $R<0$,
we shall hypothesize instead that that the spacetime has the structure of an anti--deSitter space $AdS_4$.

Unfortunately, G{\"u}rsey could not realize that the decomposition he proposed is due to a spontaneous breakdown
of conformal symmetry, since this kind of concept was still in embryo  in those years \cite{NAMBU} \cite{BAKER}.
Thus, the question arises as to whether his view is entirely compatible with the assumption of a conformal symmetry
breakdown.

To overcome these inconveniences, we propose a different approach. At variance with G{\"u}rsey, we assume that the scale
factor $e^{\alpha(x)}$ of the expansion of the universe on the large scale is determined by the motion equation of
the dilation field $\sigma(x) =\sigma_0\,e^{\alpha(x)}$ interacting with a scalar field, as described in \S\,\ref{G2MneedsH},
so that $e^{\alpha(x)}$ is subject to the boundary conditions
\begin{equation}
\label{limalpha}
e^{\alpha(0)} \ll 1 \,,\quad
\lim_{\tau\rightarrow \infty} \frac{e^{\alpha(x)}}{c_R(\tau)}\rightarrow 1\,,
\end{equation}
(not granted {\em a priori}) where $c_R(\tau)$ is the acceleration factor of a homogeneous GR universe of constant
spacetime curvature $R<0$, so that CGR converges to GR for $\tau\rightarrow \infty$.

As we shall see in the end of our investigation, these conditions force $e^{\alpha(x)}$ to
acquire a sigmoid--shaped profile, as expected in any reliable theory of inflationary cosmology.

\subsection{General form of conformal geodesic equations}
\label{confgeods}
On the Cartan manifold, the motion of a point--like test particle under the
action of the conformal gravitational field is governed by the conformal geodesic
equation
\begin{equation}
\label{cartangeods}
\frac{d^2 x^\lambda}{d\hat s^2}+\hat\Gamma^\lambda_{\mu\nu}\frac{d
x^\mu}{d\hat s}\frac{dx^\nu}{d\hat s}= \frac{dx^\rho}{d\hat s}\Bigl(\hat D_\rho \frac{dx^\lambda}{d\hat
s}\Bigr)=0 \,\,\,\, \hbox{(self--parallelism condition)}\,,
\end{equation}
where  $d\hat s= \sqrt{\hat g_{\mu\nu}dx^\mu dx^\nu}=(\sigma/\sigma_0)\, ds= e^{\alpha}ds$
is the proper--time element of the particle along its geode\-sic, $\hat D_\rho$ the covariant
derivatives on the Cartan manifold and
\begin{equation}
\label{tildeGammas2Gammas}
\hat \Gamma^\lambda_{\mu\nu} =
\delta^\lambda_\mu\partial_\nu\alpha + \delta^\lambda_\nu\partial_\mu\alpha -
g_{\mu\nu}\partial^\lambda \alpha + \Gamma^\lambda_{\mu\nu}\,, \,\, \hbox{with } \Gamma^\lambda_{\mu\nu}=
\frac{1}{2} g^{\rho\lambda}\bigl(\partial_\mu g_{\rho\nu} + \partial_\nu g_{\rho\mu}-\partial_\rho g_{\mu\nu}
\bigr)\,,
\end{equation}
are the Christoffel symbols constructed out of $\tilde g_{\mu\nu}$. Multiplying Eq(\ref{cartangeods}) by $e^{2\alpha}$,
using Eq (\ref{tildeGammas2Gammas}) and putting $d\tilde s = e^\alpha ds$, we
obtain the geodesic equation on the Riemann manifold
\begin{equation}
\label{riemgeods}
e^{\alpha}\frac{d}{ds}\Bigl(e^{-\alpha}\frac{d x^\lambda}{ds}\Bigr) +
2\,\big(\partial_\mu\alpha\big)\,\frac{dx^\mu}{ds} \frac{d x^\lambda}{ds}-
\partial^\lambda\alpha+ g^{\lambda\rho}\Bigl(\partial_\mu g_{\rho\nu} -
\frac{1}{2}\partial_\rho g_{\mu\nu}\Bigr)\frac{d x^\mu}{ds}\frac{dx^\nu}{ds}=0\,,
\end{equation}
from which we obtain the contravariant 4D--acceleration of the test particle
\begin{equation}
\label{4Daccel}
a^\lambda\equiv \frac{d^2 x^\lambda}{ds^2} =
\partial^\lambda\alpha -\frac{d\alpha}{ds} u^\lambda - g^{\lambda\rho}\Bigl(\partial_\mu g_{\rho\nu}-
\frac{1}{2}\partial_\rho g_{\mu\nu}\Bigr)u^\mu u^\nu\,,
\end{equation}
where $u^\mu = dx^\mu/ds$ is the contravariant 4D--velocity vector.

The first two terms on the right--hand side represent the contribution to acceleration due to the
dilation field and the last term the contribution due to the gravitational field.

It is thus clear that the dilation field also exerts an inertial force, which disappears as
$\alpha$ approaches zero at the end of the inflationary epoch. Thus, in the post--inflationary
era, the 4D acceleration of the test particle tends to depend only on the gravitational field
and, therefore, CGR tends to behave like GR.

\subsection{Hyperbolic polar coordinates and synchronized comoving observers}
\label{synchrobs}
All world--lines stemming from a point $V$ of a smooth Riemann manifold and propagating within the future cone of origin
$V$ will be called the {\em polar geodesics} from $V$. By means of a suitable diffeomorphism of the manifold, we will
be able to choose the metric of the future cone near $V$ as that of a Minkowski spacetime. Let us see
how the set of all polar geodesics stemming from $V$ can be used to implement a system of {\em hyperbolic polar coordinates}.

Any polar geodesic being one--to--one with its direction $\vec\rho\,$ at $V$ can be denoted as $\Gamma(\vec\rho\,)$.
In particular, any polar geodesic, but in general only one -- say $\Gamma(\vec\rho_0)\equiv \Gamma(0)$ -- can be transformed
into a straight axis by a second diffeomorphism of the manifold, without altering the metric near $V$. Let us define the
{\em kinematic time} $\tau$ of an event $O\in\Gamma(\vec\rho\,)$ as the length of the geodesic segment $VO$, and then
the {\em hyperbolic angle} $\varrho\,$ as the derivative with respect to $\tau$, at $\tau=0$, of the length of the oriented
hyperbolic arc $\vec\rho\,$ between $\Gamma(0)$ and $\Gamma(\vec\rho\,)$, as shown in Fig.\,2. Lastly, let us indicate by
$\{\theta, \phi\}$ the Euler angles of the projection $\vec r$ of $\Gamma(\vec\rho\,)$ onto the 3D--plane orthogonal
to $\Gamma(0)$ at $V$. Since the metric in the neighborhood of $V$ is Minkowskian, we are in a position to put
$\vec\rho=\{\varrho, \theta, \phi\}$ and $\vec\rho_0=\{0, 0, 0\}$.
\begin{figure}[!ht]
\centering
\mbox{%
\begin{minipage}{0.4\textwidth}
\includegraphics[scale=0.8]{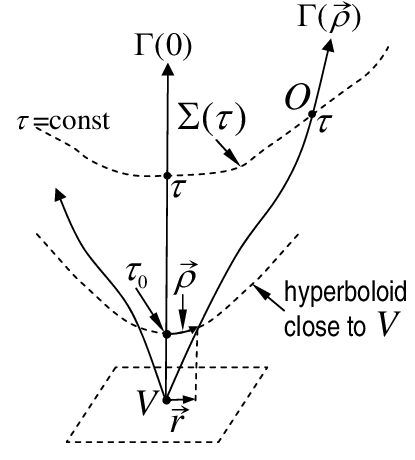}
\end{minipage}%
\quad
\begin{minipage}[c]{0.53\textwidth}
\caption{\small  Geodesics passing through a point $V$ of a spacetime manifold and spanning
the interior of the future cone of origin $V$ can be parameterized by {\em hyperbolic polar coordinates}
$\{\tau, \vec\rho\,\}$. This is possible because each geodesic $\Gamma(\vec\rho\,)$ depends uniquely on its
direction $\vec\rho\,=\{\varrho, \theta,\phi\}$ at $V$. We can then define the time $\tau$ of an event
$O\in\Gamma(\vec\rho\,)$ as the length of the geodesic segment $VO$. The 3D surface $\Sigma(\tau)$
is the locus of all the events which have the same time  $\tau$.}
\end{minipage}
}
\end{figure}

Since each line--element $ds$ of a polar geodesic has length $d\tau$ and
$\vec\rho= \{\varrho, \theta,\phi\}\equiv \{\rho^1, \rho^2,\rho^3\}$ is a constant triplet, we have $d\tau/ds =1$
and $d\rho^i/ds =0$, we can then cast any squared line--element lying in the future cone in the form
\begin{equation}
\label{pollinelement} ds^2 = d\tau^2 -\tau^2 \,\gamma_{ij}(\tau, \vec\rho\,)\,d\rho^i d\rho^j\,,\quad (i, j
=1,2,3)\,,
\end{equation}
with initial conditions
\begin{eqnarray}
\lim_{\tau \rightarrow 0}\gamma_{11} = 1;\,\,\, \lim_{\tau \rightarrow 0}\gamma_{22}=(\sinh\varrho)^2;\,\,\,
\lim_{\tau \rightarrow 0}\gamma_{33} =(\sinh\varrho\,\sin\vartheta)^2;\,\,\,
\lim_{\tau \rightarrow 0}\gamma_{ij}=0\,\, (i\neq j).\nonumber
\end{eqnarray}

This shows that the volume element of the interior of a future cone can be expressed as
$\sqrt{-g(x)}\,d^4x = \sqrt{-\gamma(\tau, \vec\rho\,)}\,d\tau d^3\rho$, where $d^3\rho\equiv d\rho^1d\rho^2d\rho^3
\equiv d\varrho\,d\theta\,d\phi$.

If the metric is not so curved as to require a multi--chart representation, the information about the
gravitational field is completely incorporated into coefficients $\gamma_{ij}(\tau, \vec\rho\,)$.

Observers moving along the polar geodesics stemming from $V$ and using $\tau$ as the measure of time
are called synchronized and comoving. Hence, $\Sigma(\tau)$ represents the set of synchronized
comoving observers at time $\tau$.

The set of all events which have the same time $\tau$, for all polar geodesics of the
future cone, forms a 3D subspace $\Sigma(\tau)$. A point $O$, running along one of these geodesics,
is presumed to represent an observer on the Riemann manifold whose clock marks $\tau$.
Since the observers of a universe expanding in a future cone of origin $V$ are called ``comoving",
provided that they move along their own polar geodesics stemming from $V$, we can say that
$\Sigma(\tau)$ represents the set of synchronized comoving observers at kinematic time $\tau$.

An important point regarding this internal ordering of a future cone is that the 3D volumes of $\Sigma(\tau)$
is infinite at any $\tau$, in contrast with those of the 3D sections orthogonal to the time axis of a
future cone in the standard representation of GR spacetime. This allows us to
define the {\em thermodynamic limit} of the universe in all 3D space of comoving observers,
which is essential for describing the irreversible dispersion of infrared photons to infinity
and the macroscopic evolution of the universe as a thermodynamic process.

The simplest example of hyperbolic polar coordinates is given by expressing $\{\tau,\vec \rho\,\}$
as functions of Lorentzian coordinates $x^\mu = \{x^0, x^1, x^2, x^3\}$, via $x^0 = \tau\cosh\rho$,
$x^1 = \tau\sinh\varrho \sin\vartheta\cos\phi$, $x^2 = \tau\sinh\varrho\sin\vartheta\sin\phi$, $x^3 =
\tau\sinh\varrho\cos\vartheta$, from which we derive $r \equiv \sqrt{(x^1)^2+(x^2)^2+(x^3)^2}=\tau\sinh\varrho$,
$\tau  =  \sqrt{(x^0)^2 - (x^1)^2-(x^2)^2-(x^3)^2}$, as shown in Fig.\,2.
\begin{figure}[h]
\centering
\mbox{%
\begin{minipage}{.4\textwidth}
\includegraphics[scale=0.45]{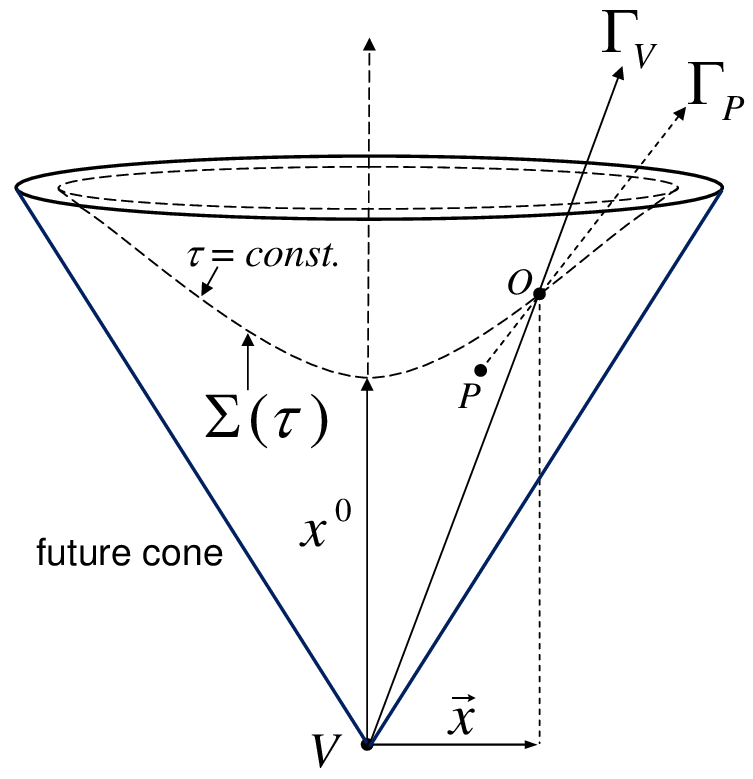}
\end{minipage}%
\quad
\begin{minipage}[c]{.5\textwidth}
\caption{\small Lorentzian coordinates spanning a future cone in Minkowski spacetime.
Synchronized comoving observers lie on the 3D hyperboloid $\Sigma(\tau)$ at kinematic--time
distance $x^0=\tau$ from vertex $V$. The position of an observer O on $\Sigma(\tau)$ is identified
by the radial vector $\vec x$.}
\end{minipage}%
}
\end{figure}

The reader can easily verify that in hyperbolic coordinates the expression of the squared line element
$ds^2 = (dx^0)^2 - (dx^1)^2-(dx^2)^2-(dx^3)^2$ takes the form
$$ds^2 = d\tau^2 - \tau^2\bigl[d\varrho^2+ (\sinh \varrho)^2 d\vartheta^2 +
(\sinh \varrho \sin \vartheta)^2 d\phi^2\bigr].$$

If the gravitational forces are negligible and the matter distribution on the large scale
is homogeneous and isotropic all over each $\Sigma(\tau)$, we can assume that the
expansion factor $e^{\alpha(x)}$ is the same all over $\Sigma(\tau)$, i.e., $\alpha(x)\equiv \alpha(\tau)$.
Since in this case the Riemann manifold is conformally flat, we can parameterize the future cone by Lorentzian
coordinates $\{x^0, \vec x\}\equiv \{x^0, x^1, x^2, x^3\}$, with metric $\eta_{\mu\nu}=\hbox{diag}(1, -1, -1, -1)$.

We can thus express kinematic time as $\tau=\sqrt{(x^0)^2- \vert\vec x \vert^2}$,
and therefore envisage $\vec{v}_O= -\vec x/x^0$ as the contravariant velocity vector
of comoving observer $O$ at $\{x^0, \vec x\}$
and $v_O= \vert\vec{v}_O\vert$ its norm. The following relationships are then easily proven
\begin{equation}
\label{velobs}
\frac{d\tau}{dx^0} =\frac{1}{\sqrt{1-v_O^2}}\,; \quad {\vec
\nabla}\tau = \frac{\vec{v}_O}{\sqrt{1-v_O^2}}\,,
\end{equation}
where $\vec \nabla= \{\partial^1, \partial^2, \partial^3 \}$ and $\partial^i \equiv \eta^{ij}\partial_j = -
\partial_i\,,\,\, (i,j = 1,2,3)$.

Now, let us consider a test particle $P$ moving along a geodesic $\Gamma_P$, generally
not stemming from $V$, let $s_P$ be its proper time and
$x^\mu_P$ its spacetime coordinates. Hence, $ds_P$  is related to $dx^\mu_P$
by $ds^2_P= g_{\mu\nu}\,dx^\mu_P dx^\nu_P$ and the contravariant
components $u^\mu_P$ of 4D--velocity  are
$$
u^0_P = \frac{d x^0}{ds_P}= \frac{1}{\sqrt{1-v_P^2}}\,,\quad \vec u_P = \frac{d \vec x_P}{ds_P}=
\frac{d \vec x^P}{dx_0^P} \frac{d x^0_P}{ds_P}= \frac{\vec v_P}{\sqrt{1-v^2_P}}\,,
$$
where $\vec v_P$ is the 3D-velocity of the particle at $x$. Since, by hypothesis,
$\alpha$ depends only on $\tau$ and $g_{\mu\nu}=\eta_{\mu\nu}$, Eq.(\ref{4Daccel}) becomes
$$
a^\lambda = \partial^\lambda \alpha -\frac{d\alpha}{ds} u^\lambda
= \frac{d\alpha}{d\tau}\Bigl(\partial^\lambda\tau -\frac{d\tau}{ds} u^\lambda\Bigr) \,.
$$
Using Eqs.(\ref{velobs}) and
$$
\frac{d\tau}{ds} = \frac{d\tau}{dx^0 }\frac{dx^0}{ds} = \frac{\sqrt{1-v^2_P}}{\sqrt{1-v^2_O}} \,,
$$
we obtain the 3D--acceleration
\begin{equation}
\label{3accs}
\vec a_P =\frac{\vec v_O -\vec v_P }{\sqrt{1-v^2_O}} \frac{d\alpha}{d\tau}\,,
\end{equation}
showing that the test particle is not only subject to the gravitational forces described by the last term on the
right--hand  member of Eq (\ref{4Daccel}), but also to a viscous force
proportional to the slope of the dilation field profile, which makes $\vec v_P$ converge to $\vec v_O$.

This suggests regarding the dilation field in expansion as a viscous medium which keeps matter in the reference frame
of comoving synchronized observers, as the privileged reference frame of fixed stars invoked by the Mach principle
is expected to do.

\renewcommand{\theequation}{A-\arabic{equation}}
\setcounter{equation}{0}  % reset counter
\addcontentsline{toc}{section}{Appendix - Basic formulas of standard-- and conformal--tensor calculus}
\section*{\qquad\qquad\qquad\qquad\qquad  APPENDIX\\
Basic formulas of  standard-- and conformal--tensor calculus}
\label{basform}
For the sake of clarity, and for the purpose of indicating a few sign conventions, we list here the
basic formulae of tensor calculus used here. For simplicity, and because
the subject is rarely mentioned in this paper, we avoid dealing with spinors,
vierbeins and their covariant derivatives. Let us start from the familiar standard--tensor calculus
of GR:
\begin{itemize}
\item[-] {\it Christoffel symbols and their metric--tensor variations:}
\begin{eqnarray}
\label{Christoff}
& & \hspace{-8mm}\Gamma_{\mu\nu}^\lambda =  \frac{1}{2} g^{\rho\lambda}\bigl(
\partial_\mu g_{\rho\nu} + \partial_\nu g_{\rho\mu}-\partial_\rho g_{\mu\nu}
\bigr)\,; \\
& & \hspace{-8mm}\delta\Gamma_{\mu\nu}^\lambda = \frac{1}{2} g^{\rho\lambda}\bigl( D_\mu
\delta g_{\rho\nu} + D_\nu \delta g_{\rho\mu}-D_\rho \delta g_{\mu\nu}
\bigr)\,; \nonumber
\end{eqnarray}
where $\delta g_{\mu\nu}(x)$ are small arbitrary variations of $g_{\mu\nu}(x)$ and $D_\mu \delta g_{\nu\lambda} \equiv \partial_\mu \delta g_{\nu\lambda}-\Gamma_{\mu\nu}^\rho \delta g_{\rho\lambda} - \Gamma_{\mu\lambda}^\rho  \delta g_{\nu\rho}$ are the
covariant derivatives of $\delta g_{\nu\lambda}$ as functions of $\Gamma_{\mu\nu}^\lambda$.

For diagonal metrics $[g_{\mu\nu}] = \mbox{diag}[h_0, h_1, \dots, h_{n-1}]$, Eqs.(\ref{Christoff}) simplify to
\begin{eqnarray}
\label{Nonzerogammas}
\Gamma^\rho_{\mu\nu} & = & 0\,\,\, (\rho, \mu, \nu \neq)\,,
\quad \Gamma^\rho_{\mu\mu} = -\frac{\partial_\rho h_\mu}{2h_\rho}\,\,\,(\rho\neq \mu)\,,\nonumber \\
\Gamma^{\rho}_{\rho\nu} & = & \frac{\partial_{\nu} h_{\rho}}{2\,h_{\rho}}\,\,\, (\rho \neq\nu)\,,\quad
\Gamma^{\rho}_{\rho \rho} = \frac{\partial_{\rho} h_{\rho}}{2\,h_{\rho}}\,,
\end{eqnarray}
where repeated indices are not to be summed.

\item[-] {\em Covariant and contravariant derivatives of mixed tensors $T^{\sigma\dots}_{\quad\,\lambda\dots}$}:
\begin{eqnarray}
\label{Dvmu}
D_\mu T^{\sigma\dots}_{\quad\,\lambda\dots} & = & \partial_\mu T^{\sigma\dots}_{\quad\,\lambda\dots}
+ \Gamma_{\mu\rho}^\sigma T^{\rho\dots}_{\quad\,\lambda\dots} + \dots - \Gamma_{\mu\lambda}^\rho
T^{\sigma\dots}_{\quad\,\rho\dots} - \dots\\
\label{contraDvmu}
D^\mu T^{\sigma\dots}_{\quad\,\lambda\dots} & = & \partial^{\mu}T^{\sigma\dots}_{\quad\,\lambda\dots}
+ \Gamma_{\rho}^{\mu\sigma} T^{\rho\dots}_{\quad\,\lambda\dots} + \dots - \Gamma_{\lambda}^{\mu\rho}
T^{\sigma\dots}_{\quad\,\rho\dots} - \dots
\end{eqnarray}
{\em where $ \Gamma_{\rho}^{\mu\sigma} \equiv g^{\mu\nu}\Gamma_{\nu\rho}^\sigma $}.
Since the following identities hold:
\vspace{-1mm}
\begin{eqnarray}
\label{Dmugdownmunu} & & D_\mu g_{\nu\lambda} \equiv \partial_\mu g_{\nu\lambda}-
\Gamma_{\mu\nu}^\rho g_{\rho\lambda} - \Gamma_{\mu\lambda}^\rho g_{\nu\rho}= 0\,;\\
\vspace{-1mm}
\label{Dmugupmunu} & & D_\mu g^{\sigma\lambda} \equiv \partial_\mu g^{\sigma\lambda}+ \Gamma_{\mu\rho}^\sigma
g^{\rho\lambda} + \Gamma_{\mu\rho}^\lambda g^{\sigma\rho}= 0\,;
\vspace{-1mm}
\end{eqnarray}
we have $D_\mu \bigl(g_{\nu\lambda}\,T^{\sigma\dots}_{\quad\,\lambda\dots}\bigr) =
g_{\nu\lambda} D_\mu T^{\sigma\dots}_{\quad\,\lambda\dots}$ and $D_\mu \bigl(g^{\nu\lambda}\,
T^{\sigma\dots}_{\quad\,\lambda\dots}\bigr) = g^{\nu\lambda} D_\mu T^{\sigma\dots}_{\quad\,\lambda\dots}$
for any tensor $T^{\sigma\dots}_{\quad\,\lambda\dots}$. Since $D^\mu\dots  =
g^{\mu\nu}D_\nu \dots =D_\nu g^{\mu\nu}\dots$, the same property holds for contravariant derivatives.
In short, both $g_{\mu\nu}$ and $g^{\mu\nu}$ and any function of these are ``transparent'' to
covariant derivatives. In particular, $D_\mu (\sqrt{-g}\,T^{\sigma\dots}_{\quad\,\lambda\dots})=
\sqrt{-g}\, D_\mu T^{\sigma\dots}_{\quad\,\lambda\dots}$, where $g$ is the determinant of matrix
$\bigl[g_{\mu\nu}\bigr]$.

\item[-] {\em Covariant divergence of a tensor $T^{\mu\nu\dots\rho}$
and contravariant divergence of a tensor $T^\mu_{\,\cdot\,\nu\dots\rho} \equiv
g^{\mu\sigma}T_{\sigma\nu\dots\rho}$ reed as follows}
\vspace{-1mm}
\begin{eqnarray}
\label{covdiv} \hspace{-5mm} D_\mu T^{\mu\nu\dots\rho} &\!\! =\!\! &
\frac{1}{\sqrt{-g}}\,\partial_\mu \big(\sqrt{-g}\, T^{\mu\nu\dots\rho}\big)
+ \Gamma_{\mu\lambda}^\nu T^{\mu\lambda\dots\rho} + \dots + \Gamma_{\mu\lambda}^\rho
T^{\mu\nu\dots\lambda};\nonumber\\
\vspace{-1mm}
D^\mu T_{\mu\nu\dots\rho} &\!\! = \!\! & \frac{1}{\sqrt{-g}}
\partial^\mu \big(\sqrt{-g}\, T_{\mu\nu\dots\rho}\big)
-\Gamma_{\nu}^{\lambda\mu} T_{\mu\lambda\dots\rho} - \dots -
\Gamma_{\mu}^{\lambda\mu} T_{\mu\nu\dots\lambda}\,;\nonumber\\
\vspace{-1mm}
D_\mu T^{\mu}_{\,\,\cdot\,\nu\dots\lambda} & = &  \frac{1}{\sqrt{-g}}\,\partial_\mu \big(\sqrt{-g}\,
T^{\mu}_{\,\,\cdot\,\nu\dots\lambda}\big)  - \Gamma_{\mu\nu}^\rho T^{\mu}_{\,\,\cdot\,\rho\dots\lambda} - \dots-
\Gamma_{\mu\lambda}^\rho T^{\mu}_{\,\,\cdot\,\nu\dots\rho}\,.
\vspace{-1mm}
\end{eqnarray}

Therefore, for consistency with Eqs (\ref{Dvmu}), (\ref{contraDvmu}) and (\ref{Dmugupmunu}), we immediately obtain
$$
\frac{1}{\sqrt{-g}}\,\partial_\mu \sqrt{-g}= \Gamma^\lambda_{\mu\lambda}\,;\quad \frac{1}{\sqrt{-g}}\,\partial_\mu
\big(\sqrt{-g}\,g^{\mu\nu}\big) =\Gamma^\nu_{\rho\sigma}g^{\rho\sigma}\,.
$$

\item[-] {\em Riemann tensor and its variations:}
\begin{eqnarray}
& & \hspace{-10mm}R^\rho_{\, .\,\mu\sigma\nu}  = \partial_\sigma \Gamma^\rho_{\nu\mu}
-\partial_\nu \Gamma^\rho_{\mu\sigma}+
\Gamma^\lambda_{\mu\nu}\Gamma^\rho_{\lambda\sigma}-
\Gamma^\lambda_{\mu\sigma}\Gamma^\rho_{\lambda\nu}\,; \nonumber\\
& & \hspace{-10mm}\delta R^\rho_{\, .\,\mu\sigma\nu}  =
\frac{1}{2}\,g^{\rho\lambda}\big( D_\sigma D_\mu \delta g_{\nu\lambda} + D_\sigma
D_\nu\delta g_{\lambda\mu}-D_\sigma D_\lambda \delta g_{\nu\mu} +\nonumber \\
& & \qquad\quad  D_\nu D_\lambda\delta g_{\mu\sigma}-D_\nu D_\mu\delta g_{\lambda\sigma}-
D_\nu D_\sigma\delta g_{\lambda\mu} \big)\,;\nonumber
\end{eqnarray}
where $\delta R^\rho_{\, .\,\mu\sigma\nu}$ are the variations of $R^\rho_{\, .\,\mu\sigma\nu}$ caused
by metric--tensor variations $\delta g_{\nu\lambda}$.

For diagonal metrics $[g_{\mu\nu}] = \mbox{diag}[h_0, h_1, \dots, h_{n-1}]$ we have (Eisenhart 1949, p.44)
\begin{eqnarray}
& & \hspace{-6mm}R_{\rho\mu\sigma\nu}=0 \quad (\rho,\mu,\sigma, \nu \neq)\,;\nonumber\\
& & \hspace{-6mm}R_{\rho\mu\mu\nu}= |h_\mu|^\frac{1}{2}\Big[\partial_\rho\partial_\nu |h_\mu|^\frac{1}{2}-
\bigl(\partial_\rho |h_\mu|^\frac{1}{2}\bigr)\partial_\nu \ln |h_\rho|^\frac{1}{2}-
\bigl(\partial_\nu |h_\mu|^\frac{1}{2}\bigr)\partial_\rho \ln |h_\nu|^\frac{1}{2}\Big]\,\, (\rho, \mu, \nu \neq);
\nonumber\\
& & \hspace{-6mm}R_{\rho\mu\mu\rho} = |h_\rho h_\mu|^\frac{1}{2}\bigg[\partial_\rho
\bigg(\frac{\partial_\rho |h_\mu|^\frac{1}{2}}{|h_\rho|^{\frac{1}{2}}}\bigg) +
\partial_\mu\bigg(\frac{\partial_\mu |h_\rho|^\frac{1}{2}}{|h_\mu|^{\frac{1}{2}}}\bigg)+
{\sum}'_\lambda \bigg(\frac{\partial_\lambda |h_\rho|^\frac{1}{2}}{|h_\lambda|^{\frac{1}{2}}}\bigg)
\partial_\lambda |h_\mu|^\frac{1}{2}\bigg]\,\, (\rho\neq\mu);\nonumber
\end{eqnarray}
where ${\sum}'_\mu$ indicates the sum for $\lambda = 0, 1,\dots , n-1$ excluding $\lambda = \mu$ and $\lambda = \rho$.

\item[-] {\em Covariant derivative commutators on covariant vectors $v_\rho$ and scalars $f$:}
\begin{equation}
\label{covdevcomms}
(D_\mu D_\nu - D_\nu D_\mu)\, v_\rho = R^\sigma_{\,\cdot\,\rho\mu\nu}\,v_\sigma\,;\quad
(D_\mu D_\nu - D_\nu D_\mu)\,f=0\,.
\end{equation}
the second of which implies $D^2 D_\nu f =D_\nu D^2f$.

\item[-] {\em The Beltrami--d'Alembert operator on scalars and vectors:}
\begin{eqnarray}
\label{D2f}
D^2 f  & \equiv & D^\mu D_\mu f = \partial_\mu\partial^\mu f - \Gamma^{\mu}_{\mu\rho} \partial^\rho f=
\frac{1}{\sqrt{-g}}\,\partial_\mu \bigl(\sqrt{-g}\,\partial^\mu f\bigr)\,;\\
\label{D2vrho}
D^2  v_\rho & \equiv & D_\mu D^\mu v_\rho = \frac{1}{\sqrt{-g}}\,
\partial_\mu \bigl(\sqrt{-g}\,\partial^\mu v_\rho\bigr)-\Gamma_{\rho\mu}^{\lambda} \partial^\mu v_\lambda\,;
\end{eqnarray}
because $\Gamma^{\nu}_{\nu\mu}(x)=\partial_\mu \ln \sqrt{-g(x)}$ as can easily be proven
with Eq.(\ref{Dmugdownmunu}) or (\ref{Dmugupmunu}) and the well--known formula $\partial_\mu \ln g(x) =
g^{\rho\sigma}(x)\,\partial_\mu g_{\rho\sigma}(x)= - g_{\rho\sigma}(x)\,\partial_\mu g^{\rho\sigma}(x)$.

\item[-] {\em Ricci tensors and their metric--tensor variations:}
\begin{eqnarray}
& & R_{\mu\nu}\equiv  R^\rho_{\, .\,\mu\rho\nu}  = \partial_\rho
\Gamma^\rho_{\mu\nu} -\partial_\nu \Gamma^\rho_{\mu\rho}+
\Gamma^\lambda_{\mu\nu}\Gamma^\rho_{\lambda\rho}-
\Gamma^\lambda_{\mu\rho}\Gamma^\rho_{\lambda\nu}\,; \nonumber\\
& & \delta R_{\mu\nu}  =  \frac{1}{2} \bigl(D^\rho D_\mu\,\delta g_{\rho\nu}
+D^\rho D_\nu\,\delta g_{\rho\mu} - D^2\, \delta g_{\mu\nu}-D_\mu D_\nu
g^{\rho\sigma}\, \delta g_{\rho\sigma}\bigr)\,, \nonumber\\
& & \label{Rvariation} R \equiv R_{\mu\nu}g^{\mu\nu}\,,\quad
\delta R = R_{\mu\nu}\, \delta g^{\mu\nu}+\frac{1}{2}\bigl(g_{\mu\nu} D^2-D_\mu D_\nu\bigr)\, \delta
g^{\mu\nu}\,;
\end{eqnarray}
The sign convention for the Riemann tensor is that of Eisenhart, but that of the Ricci tensors
is opposite to Eisenhart's, which is $R_{\mu\nu}\equiv R^\rho_{\, .\,\mu\nu\rho}= -R^\rho_{\, .\,\mu\rho\nu}$,
and matches Landau--Lifchitz (1970). The last of Eqs.(\ref{Rvariation}) yields the useful formula:
\begin{equation}
\label{useful formula}
\frac{1}{\sqrt{-g}} \frac{\delta}{\delta g^{\mu\nu}}\int \sqrt{-g}\,f\,R\,d^nx
= f\bigl(R_{\mu\nu} - \frac{1}{2}g_{\mu\nu}R\bigr)+\bigl(g_{\mu\nu}D^2 - D_\mu
D_\nu \bigr)f\,.
\end{equation}

\item[-] {\em The geometrical meaning of Ricci tensors:}
To clarify the geometrical meaning of $R_{\mu\nu}(x)$ and $R(x)$, we solve equation
$$
\big[R_{\mu\nu}(x) - c(x)\, g_{\mu\nu}(x)\big]\lambda^\mu(x) \equiv R_{\mu\nu}(x)\lambda^\mu(x)
- c(x)\,\lambda_\nu(x) = 0\,,
$$
the $n$ solutions of which, $\lambda^\mu_k(x)$, respectively associated with eigenvalues $c_k(x)$ $(k=1,2,\dots, n)$,
satisfy the orthonormalization conditions $\lambda^\mu_k(x) \lambda_{\mu h}(x) = \delta_{kh}$.
We can then write $R_{\mu\nu}(x) = \sum_k c_k(x) \lambda_{\mu k}(x)\lambda_{\nu k}(x)$  and interpret $c_k(x)$ as the
spacetime curvature at $x$ along the {\em principal direction} $\lambda^\mu_k(x)$. The interesting
formula $R(x) = \sum_k c_k(x)$ then follows. Since curvatures at $x$ may conspire to
make $\sum_k c_k(x)=0$, we see that $R(x)=0$ does not imply $R_{\mu\nu}(x)=0$. However,
since spacetime curvatures in general change from point to point, it is very probable that this
happens only on zero--measure sets of spacetime points. If $c_k(x)= \rho(x)$ for all $k$, we have
$R_{\mu\nu}(x)= c(x)\, g_{\mu\nu}(x)$, in which case the Ricci tensor is called {\em isotropic}.
If $c_k$ do not depend on $x$,  we have $R_{\mu\nu}(x)= \sum_k c_k \lambda_{\mu k}(x)\lambda_{\nu k}(x)$,
in which case the Ricci tensor is called {\em homogeneous}. In $n$D, the homogeneous isotropic
Ricci tensor has the form
\begin{equation}
\label{homisoriccitensors}
R_{\mu\nu}(x)= \frac{R}{n}\, g_{\mu\nu}(x)\,,
\end{equation}
where $R$ is constant and the Riemann tensor has the form
\begin{equation}
\label{constcurv}
R_{\mu\nu\rho\sigma}(x) = \frac{R}{n(n-1)}\big[ g_{\mu\rho}(x)g_{\nu\sigma}(x)
- g_{\mu\sigma}(x)g_{\nu\rho}(x)\big]\,.
\end{equation}
(Eisenhart, pp. 83, 203). Interesting theorems on $n$D spaces of constant curvature are mentioned
at the end of the Appendix.
\end{itemize}

The conformal--tensor calculus of CGR is enriched by new properties, which are obtained
by Weyl transformations of a few basic quantities. The following ones are of
decisive importance for our investigation. Taking as fundamental--tensor variation the finite
transformation $g_{\mu\nu}(x)\rightarrow \hat g_{\mu\nu}(x)=e^{2\alpha(x)}g_{\mu\nu}(x)$,
we obtain the transformations
\begin{eqnarray}
\label{Gammavariations}
&& \hspace{-14mm}\Gamma^\lambda_{\mu\nu} \rightarrow \hat \Gamma^\lambda_{\mu\nu}=
\Gamma^\lambda_{\mu\nu} + \delta^\lambda_\nu \partial_\mu \alpha +
\delta^\lambda_\mu \partial_\nu \alpha -g_{\mu\nu}\partial^\lambda\alpha \,; \\
\label{tildeRiemann}
&& \hspace{-14mm}R_{\mu\rho\sigma\nu} \rightarrow \hat R_{\mu\rho\sigma\nu} =  e^{2\alpha}\big[R_{\mu\rho\sigma\nu}
+g_{\mu\nu} A_{\rho\sigma}+g_{\rho\sigma} A_{\mu\nu}-g_{\mu\sigma}A_{\rho\nu}-g_{\rho\nu}A_{\mu\sigma} + \nonumber\\
&&\qquad(g_{\mu\nu} g_{\rho\sigma} -g_{\mu\sigma} g_{\rho\nu})(\partial^\lambda\alpha)
\partial_\lambda \alpha\big],\,\hbox{where } A_{\mu\nu}\!=\!D_\mu\partial_\nu\alpha -
(\partial_\mu\alpha)\partial_\nu\alpha;\\
\label{RmunutotildeRmu}
&& \hspace{-14mm}R_{\mu\nu}\rightarrow \hat R_{\mu\nu}= R_{\mu\nu}\!
-\!(n\!-\!2)\big[D_\mu\partial_\nu\alpha\!-\!(\partial_\mu \alpha)\partial_\nu \alpha\big]\! -\!
g_{\mu\nu}\big[D^2\alpha+ (n\!-\!2)(\partial^\rho \alpha)\partial_\rho \alpha\big]\!;\\
%\nonumber \\&& \qquad\qquad g_{\mu\nu}\big[D^2\alpha+ (n-2)(\partial^\rho \alpha)\partial_\rho \alpha\big]\,;\\
\label{RtotildeR}
&& \hspace{-14mm}R \rightarrow \hat R = e^{-2\alpha}\bigl[R
-2(n\!-\!1) D^2\alpha -(n\!-\!1)(n\!-\!2)(\partial^\rho \alpha)\partial_\rho \alpha\bigr];
\end{eqnarray}
where $\delta_\mu^\nu$ is the Kronecker delta (from Eisenhart's text, 1949, pp.89--90,
but with opposite sign convention for $R_{\mu\nu}$ and $R$). So, the conformal counterpart
$\hat G_{\mu\nu}\equiv \hat R_{\mu\nu} - \frac{1}{2}\hat g_{\mu\nu}\hat R$
of Einstein's tensor $G_{\mu\nu}\equiv R_{\mu\nu} - \frac{1}{2}g_{\mu\nu}R$ in $n$D is
\begin{equation}
\label{GmunutotildeGmunu}
\hat G_{\mu\nu} =G_{\mu\nu} - (n-2)\big[D_\mu\partial_\nu\alpha - (\partial_\mu\alpha)\partial_\nu\alpha\big]
+g_{\mu\nu}(n-2)\Big[D^2\alpha +\frac{n-3}{2}(\partial^\rho \alpha)\partial_\rho\alpha\Big].
\end{equation}

With the identities
\begin{eqnarray}
& & \!\!D_\mu\partial_\nu \alpha = e^{-\alpha} D_\mu\partial_\nu e^{\alpha} -
e^{-2\alpha} (\partial_\mu e^\alpha)\,\partial_\nu e^\alpha=  e^{-2\alpha}\big[ D_\mu(e^{\alpha}\partial_\nu e^{\alpha}) -
2\,(\partial_\mu e^\alpha)\,\partial_\nu e^\alpha\big],\nonumber\\
& & \!\!D^2\alpha = e^{-\alpha} D^2 e^{\alpha} - e^{-2\alpha}(\partial^\rho e^{\alpha})\,\partial_\rho e^{\alpha}
= e^{-2\alpha}\big[D^\rho(e^{\alpha}\partial_\rho e^{\alpha})-2\,(\partial^\rho e^{\alpha})\,
\partial_\rho e^{\alpha}\big],\nonumber
\end{eqnarray}
Eqs (\ref{RmunutotildeRmu}) (\ref{RtotildeR}) (\ref{GmunutotildeGmunu}) can be respectively cast in the forms
\begin{eqnarray}
\label{RmunutotildeRmu2}
&&\hspace{-10mm}\hat R_{\mu\nu}= R_{\mu\nu}
-(n-2)\,e^{-2\alpha} \big[e^{\alpha} D_\mu\partial_\nu e^{\alpha} -
2\,(\partial_\mu e^\alpha)\,\partial_\nu e^\alpha\big]- \nonumber \\
&&\quad g_{\mu\nu}e^{-2\alpha}\big[e^{\alpha} D^2 e^{\alpha} +(n-3)(\partial^\rho e^{\alpha})
\,\partial_\rho e^{\alpha}\big] \equiv \nonumber\\
&& \quad R_{\mu\nu}-(n-2)\,e^{-2\alpha}\big[D_\mu(e^{\alpha}\partial_\nu e^{\alpha}) -
3\,(\partial_\mu e^\alpha)\,\partial_\nu e^\alpha\big] -\nonumber \\
&& \hspace{4mm} g_{\mu\nu}\, e^{-2\alpha}\big[D^\rho(e^{\alpha}\partial_\rho e^{\alpha})-
(n-4)(\partial^\rho e^\alpha)\,\partial_\rho e^\alpha\big];\\
\label{RtotildeR2}
&&\hspace{-10mm}\hat R = e^{-2\alpha}R -(n-1)\,e^{-4\alpha}\big[(n-4)(\partial^\rho
e^{\alpha})\,\partial_\rho e^{\alpha}+2\,e^{\alpha} D^2 e^{\alpha}\bigr]\equiv\nonumber\\
&&\hspace{-2mm} e^{-2\alpha}R -(n-1)\,e^{-4\alpha}\bigl[(n-6) (\partial^\rho
e^{\alpha})\,\partial_\rho e^{\alpha}+2\,D_\mu (e^{\alpha}\partial^\mu e^{\alpha})\bigr];\\
%\end{eqnarray}
%Hence, the conformal extension of the Einstein gravitational tensor is
%\begin{eqnarray}
\label{RmunutotildeRmu2}
&&\hspace{-10mm}\hat G_{\mu\nu} = G_{\mu\nu} -(n-2)\,e^{-2\alpha}\big[e^{\alpha}D_\mu\partial_\nu e^{\alpha} -
2\,(\partial_\mu e^\alpha)\,\partial_\nu e^\alpha\big] +  \nonumber \\
&& \hspace{2mm} g_{\mu\nu} (n-2)\,e^{-2\alpha}\bigg[e^{\alpha}D^2 e^{\alpha}+\frac{(n-5)}{2}
(\partial^\rho e^{\alpha})\partial_\rho e^{\alpha}\bigg]\equiv \nonumber\\
&& \hspace{2mm} G_{\mu\nu}- (n-2)\, e^{-2\alpha}\big[D_\mu(e^{\alpha}\partial_\nu e^{\alpha})-
3\,(\partial_\mu e^{\alpha})(\partial_\nu e^{\alpha})\big]+ \nonumber \\
& & \hspace{2mm} g_{\mu\nu} (n-2)\, e^{-2\alpha}\bigg[D_\rho(g^{\rho\tau}e^{\alpha}\partial_\tau e^{\alpha})+
\frac{n-7}{2}\, (\partial^\rho e^{\alpha})\,\partial_\rho e^{\alpha}\bigg] \,.
\end{eqnarray}
In particular, putting $n=4$ and $e^{\alpha(x)}= \sigma(x)/\sigma_0$, we obtain
\begin{eqnarray}
\label{TildeRmunu4}
&&\hspace{-16mm}\hat R_{\mu\nu}= R_{\mu\nu}+\sigma^{-2}\!\big[4\,(\partial_\mu \sigma)
\,\partial_\nu \sigma - g_{\mu\nu} (\partial^\rho \sigma)\,\partial_\rho \sigma\big]-
\sigma^{-1}\big(2\,D_\mu\partial_\nu\sigma+g_{\mu\nu}D^2\sigma\big);\\
\label{TildeR4}
&&\hspace{-16mm}\hat R  =  e^{-2\alpha}\bigl(R - 6\,\sigma^{-1}D^2 \sigma\bigr)
\equiv e^{-2\alpha} R + \frac{6\, e^{-4\alpha}}
{\sigma^2_0}\bigl[(\partial^\rho\sigma)\,\partial_\rho\sigma -
D_\mu(\sigma\,\partial^\mu\sigma)\bigr];\\
\label{GmunutotildeGmunu4}
&&\hspace{-16mm}\hat G_{\mu\nu}  =
R_{\mu\nu} - \frac{1}{2}g_{\mu\nu}R  + \frac{1}{\sigma^2}\big[4(\partial_\mu\sigma)\partial_\nu\sigma -
g_{\mu\nu}(\partial^\rho \sigma)\partial_\rho\sigma\big]+\frac{2}{\sigma}(g_{\mu\nu}D^2-D_\mu\partial_\nu)\,\sigma.
\end{eqnarray}

Conformal covariant and contravariant derivatives of conformal mixed tensors mimic the standard ones:
\begin{eqnarray}
\label{tildeDmutildeTheta}
\hat D_\mu \hat T^{\sigma\dots}_{\quad\,\lambda\dots} & = & \hat \partial_\mu \hat T^{\sigma\dots}_{\quad\,\lambda\dots}
+ \hat \Gamma_{\mu\rho}^\sigma \hat T^{\rho\dots}_{\quad\,\lambda\dots} + \dots - \hat\Gamma_{\mu\lambda}^\rho
\hat T^{\sigma\dots}_{\quad\,\rho\dots} - \dots \\
\label{contratildeDmutildeTheta}
\hat D^\mu \hat T^{\sigma\dots}_{\quad\,\lambda\dots} & = & \hat \partial^{\mu}\hat T^{\sigma\dots}_{\quad\,\lambda\dots}
+ \hat \Gamma_{\rho}^{\mu\sigma} \hat T^{\rho\dots}_{\quad\,\lambda\dots} + \dots - \hat \Gamma_{\lambda}^{\mu\rho}
\hat T^{\sigma\dots}_{\quad\,\rho\dots} - \dots
\end{eqnarray}
with $\hat \partial_\mu = e^{-\alpha} \partial_\mu$, $\hat \partial^\mu = e^{\alpha} \partial^\mu$,
$\hat \Gamma_{\rho}^{\mu\sigma} = \hat g^{\mu\nu}\hat \Gamma_{\mu\rho}^\sigma$.
The vanishing of the fundamental--tensor covariant derivatives $\hat D_\mu\hat g_{\nu\lambda} = 0$,
and therefore the ``transparency'' properties $\hat D_\mu (\hat g_{\nu\lambda}\hat T^{\cdots}_{\cdots}) =
\hat g_{\nu\lambda} \hat D_\mu \hat T^{\cdots}_{\cdots}$, $\hat D_\mu (\hat g^{\nu\lambda}\,
\hat T^{\cdots}_{\cdots}) = \hat g^{\nu\lambda} \hat D_\mu \hat T^{\cdots}_{\cdots}$,
$\hat D_\mu (\sqrt{-\tilde g}\,\hat T^{\cdots}_{\cdots}) =\sqrt{-\hat g}\,\tilde D_\mu \hat T^{\cdots}_{\cdots}$,
still hold. In particular, the conformal covariant divergence of a conformal
covariant tensor with two indices can be written as
\begin{equation}
\label{tildeDvT2}
\hat D^\mu \hat T_{\mu\nu} = \frac{1}{\sqrt{-\hat g}}\,
\hat \partial_\mu \big(\sqrt{-\hat g}\, \hat g^{\mu\sigma}\hat T_{\sigma\nu}\big)
-\hat\Gamma_{\nu}^{\sigma\lambda} \hat T_{\sigma\lambda} =  \frac{1}{\sqrt{-\hat g}}\,
\hat \partial_\mu \big(\sqrt{-\hat g}\,\hat T^\mu_\nu\big)
-\hat\Gamma_{\nu\lambda}^{\sigma} \hat T^\lambda_{\sigma}\,.
\end{equation}

As already specified, symbols superscripted by a hat in Eqs.(\ref{Gammavariations})--(\ref{tildeDvT2})
describe the structural changes of the basic tensors of absolute differential calculus from GR to CGR.

Another important property of Weyl transformations regards the totally traceless part $C_{\mu\nu\rho\sigma}$ of Riemann tensor
$R_{\mu\nu\rho\sigma}=C_{\mu\nu\rho\sigma}\,+\,\cdots$, known as the {\em conformal--curvature tensor of Weyl},  which satisfies the equations
$$
C_{\mu\nu\rho\sigma}g^{\mu\nu} =
C_{\mu\nu\rho\sigma}g^{\mu\rho}=C_{\mu\nu\rho\sigma}g^{\mu\sigma}=
C_{\mu\nu\rho\sigma}g^{\nu\rho}=C_{\mu\nu\rho\sigma}g^{\nu\sigma}=
C_{\mu\nu\rho\sigma}g^{\rho\sigma}=0 \,.
$$
This property consists precisely of the invariance of mixed tensor
$C^\mu_{\,.\,\nu\rho\sigma}=g^{\mu\lambda}C_{\lambda\nu\rho\sigma}$ under Weyl
transformations, which may then be abbreviated to
$$
C^\mu_{\,.\,\nu\rho\sigma}(x)\rightarrow \hat C^\mu_{\,.\,\nu\rho\sigma}(x)
= C^\mu_{\,.\,\nu\rho\sigma}(x)\,.
$$
This means that square of the Weyl--tensor
$C^2(x)=C_{\mu\nu\rho\sigma}(x)\,C^{\mu\nu\rho\sigma}(x)$ undergoes the Weyl
transformation
$$
C^2(x)\rightarrow \hat C^2(x) = e^{-4\,\alpha(x)} C^2(x)\,.
$$

Consequently, the action integral
\begin{equation}
\label{weylterm} A^C_n = -\frac{\beta^2}{2}\int\sqrt{-g(x)}\,C^2(x)\,d^n x\,,
\end{equation}
where $\beta$ is a real constant, is conformal invariant for $n = 4$ only.

Regarding the possible presence of term  $A^C_4$  in the Einstein--Hilbert gravitational action--integral,
it has been proven by Stelle in 1977 \cite{STELLE} that any small real value  of $\beta$ suffices to guarantee
the renormalizability of quantum gravity, although at the price of introducing gravitational ghosts
of mass $M_G=M_{rP}/\beta$, where $M_{rP}$ is the reduced Planck mass, which work in practice as
Pauli--Villars regulators of graviton propagators.

\end{document}